\documentclass[%
preprint,
 amsmath,amssymb,
 aps,
]{revtex4-1}

\usepackage{graphicx}
\usepackage{dcolumn}
\usepackage{bm}

\usepackage{color}

\newcommand{\newtext}[1]{\textcolor{black}{#1}}

\begin{document}

\title{Disentangling the role of structure and friction in shear jamming}

\author{H. A. Vinutha}
\author{Srikanth Sastry}
\altaffiliation{Kavli Institute of Theoretical Physics, Kohn Hall, University of Santa Barbara, CA 93106-4030, USA}
 \affiliation{Jawaharlal Nehru Center for Advanced Scientific Research, Jakkur Campus, Bangalore 560064, India}
 \affiliation{%
  TIFR Center for Interdisciplinary Sciences, 21 Brundavan Colony, Narsingi, 500075 Hyderabad, India}

\begin{abstract}
{ Amorphous packings of spheres have been intensely investigated
  in order to understand the mechanical and flow behaviour of dense
  granular matter, and to explore universal aspects of the transition
  from fluid to structurally arrested or jammed states. Considerable
  attention has recently been focussed on anisotropic packings of
  frictional grains generated by shear deformation leading to
  \textbf{\textit {shear jamming}}, which occurs below the jamming
  density for isotropic packings of frictionless grains. With the aim
  of disentangling the role of shear deformation induced structures
  and friction in generating shear jamming, we study sheared
  assemblies of \textbf {\textit {frictionless}} spheres
  computationally, over a wide range of densities, extending far below
  the isotropic jamming point. We demonstrate the emergence of a
  variety of geometric features characteristic of jammed packings with
  the increase of shear strain. The average contact number and the
  distributions of contact forces suggest the presence of a threshold
  density, well below the isotropic jamming point, above which a
  qualitative change occurs in the jamming behaviour of sheared
  configurations. We show that above this threshold density, friction
  stabilizes the sheared configurations we generate. Our results thus
  reveal the emergence of geometric features characteristic of jammed
  states as a result of shear deformation alone, while friction is
  instrumental in stabilising packings over a range of densities below
  the isotropic jamming point.}

\end{abstract}



\maketitle


The transition from a fluid to a rigid, or jammed, state occurs and is
of interest in a wide variety of condensed matter, with glasses,
granular packings and colloidal suspensions being well known
examples. Understanding the transition, occurring variously when
temperature or applied stress is lowered, or the density raised,
involves interconnected changes in structure, thermodynamics, dynamics
of structural relaxation and rheology, and a unified and definitive
picture of this {\it jamming} transition, has been an actively pursued
goal with implications in diverse areas of research
\cite{liu-nagel-2010}. Random packings of frictionless
hard and soft spheres have been studied in this general context
\cite{liu-nagel-2010,bernal-mason-1960,scott-1960,torquatorev},
and in particular as an idealised or {\it reference} model for
granular materials. Much attention has been focussed on behaviour as
the jamming point, identified \cite{bernal-mason-1960,scott-1960} to
occur as a packing fraction of about $64 \%$, is approached. Although 
the density at which {\it random close packing} occurs is understood not to be unique (see, {\it e. g.},
\cite{torquatorev,liu-nagel-2010,pinaki-2010} and references therein),
many aspects of behaviour suggesting the jamming point (denoted
henceforth as $\phi_J$) to be a {\it critical} point\cite{ohern-2002},
appear robust \cite{liu-nagel-2010}. Real granular materials studied
experimentally inevitably deviate from this idealization, and how
these deviations influence their jamming behaviour has been an active
subject of recent
research\cite{makse-2008,silbert-2010,bi-2011,otsuki-2011,ciamarra-2011,shen-2012}. In
particular, jamming of frictional grains under shear deformation, or
{\it shear jamming}, has been shown to arise \cite{bi-2011} over a
range of stresses, and of densities below $\phi_J$, resulting in a
density-stress {\it phase diagram} that is substantially different
from the frictionless case \cite{claus-2009}.  An extended range of
jamming densities has also been discussed earlier in the context of
\textit{random loose packing}
\cite{bernal-mason-1960,scott-1960,onoda-1990},\newtext{protocol dependence \cite{bertrand}, and memory effects \cite{kumar}}, and specifically for
frictional packings \cite{makse-2008,silbert-2010}. Given that the
structural changes and organisation resulting from the shear
deformation, as well as friction likely play an important role in
generating shear jammed packings, elucidating the role of each of
these factors is central to understanding shear jamming. We address
this issue here, through a computational study of sheared
configurations of frictionless soft spheres.

We study a system of $N = 2000$ equal-sized frictionless spheres
interacting with a harmonic repulsive potential
\cite{ohern-2002,pinaki-2010} over a wide range of densities, from a
packing fraction of $0.26$ to $0.627$, generated (for high densities) by rapid compression
of fluid configurations or decompression of jammed
configurations. Shear deformation is applied through an athermal quasi
static procedure \cite{claus-2009}, by incrementing strain
$\gamma_{xz}$ in small steps (typically of $d\gamma = 5 \times
10^{-5}$, but as low as $d\gamma = 5 \times 10^{-12}$ in cases
indicated), followed by energy minimisation at each step. The
procedure is applied till a steady state in which the shear stress
$\sigma_{xz}$ (which remains very small) and the average number of
contacts $Z$ reach stable values. Further details concerning the
simulations and analysis are presented in Methods and Supplementary
Information (SI).
 
\begin{figure*}[h]
\includegraphics[width=7cm,height=7cm]{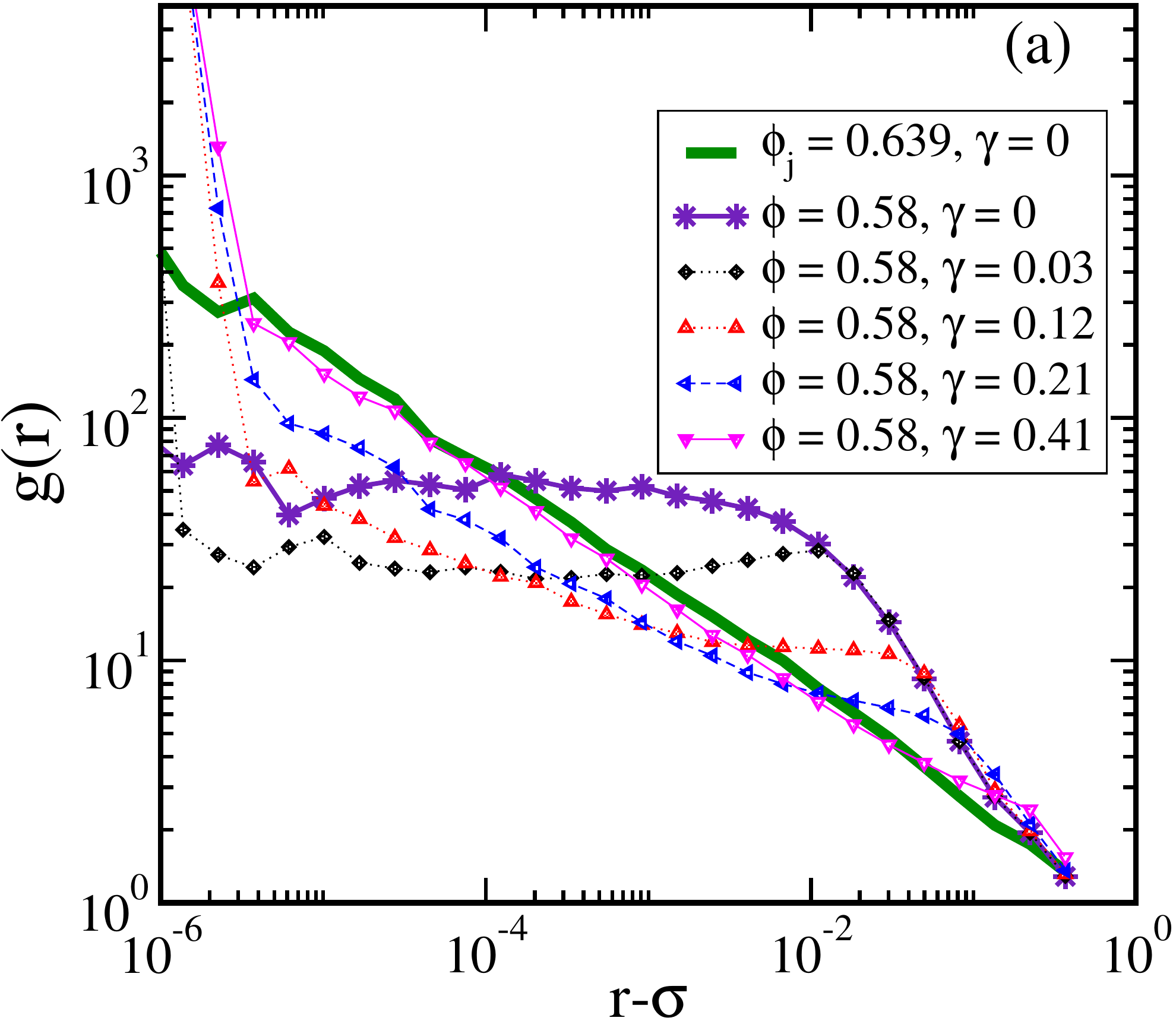}
\hskip 10mm
\includegraphics[width=7cm,height=7cm]{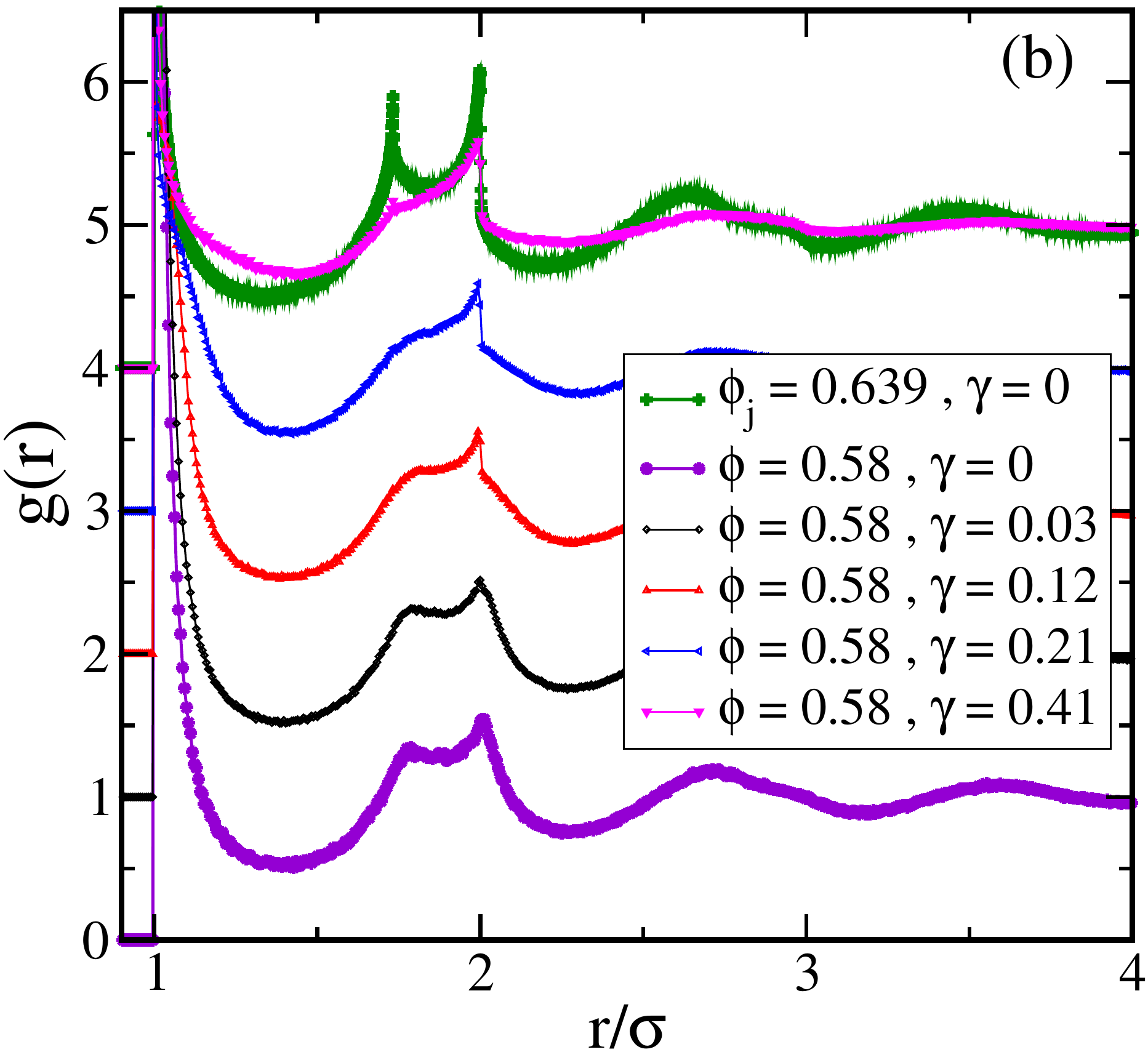}
\\
\vspace{2cm}
\includegraphics[width=7cm,height=7cm]{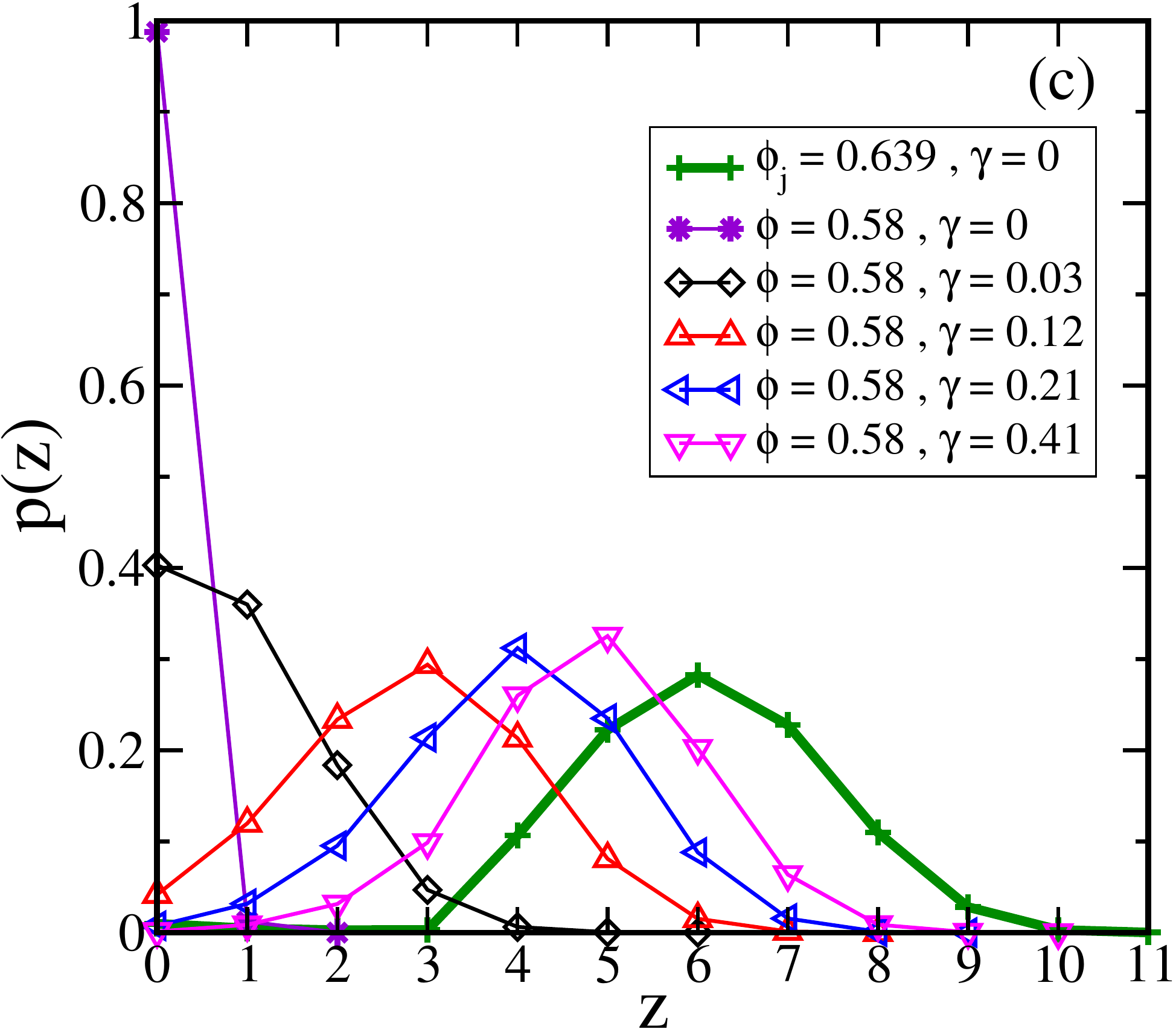}
\hskip 8mm
\includegraphics[width=7.4cm,height=7.1cm]{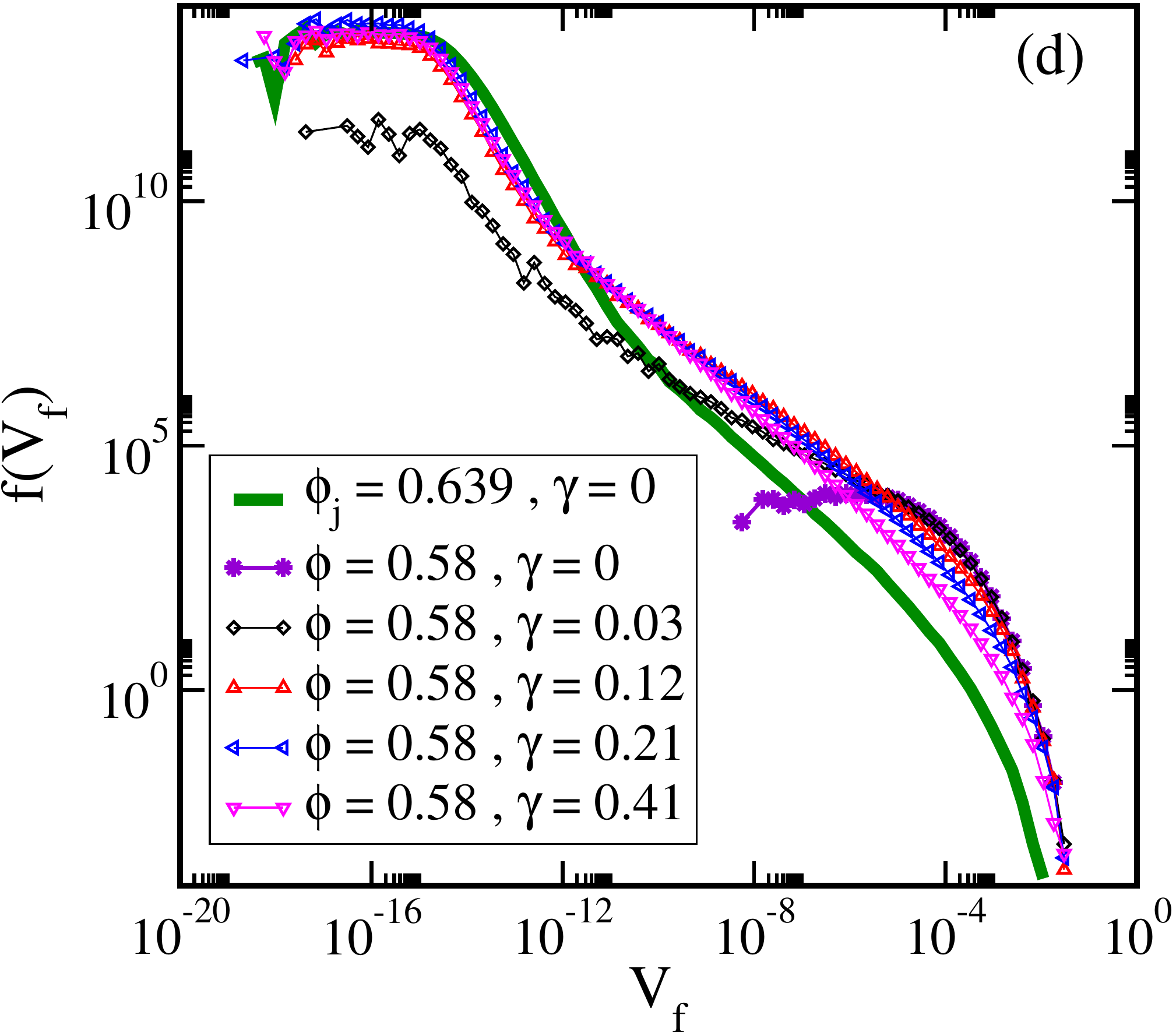}
\caption{\label{fig1} }
\end{figure*}

\begin{figure}[h]
\includegraphics[width=7cm,height=7cm]{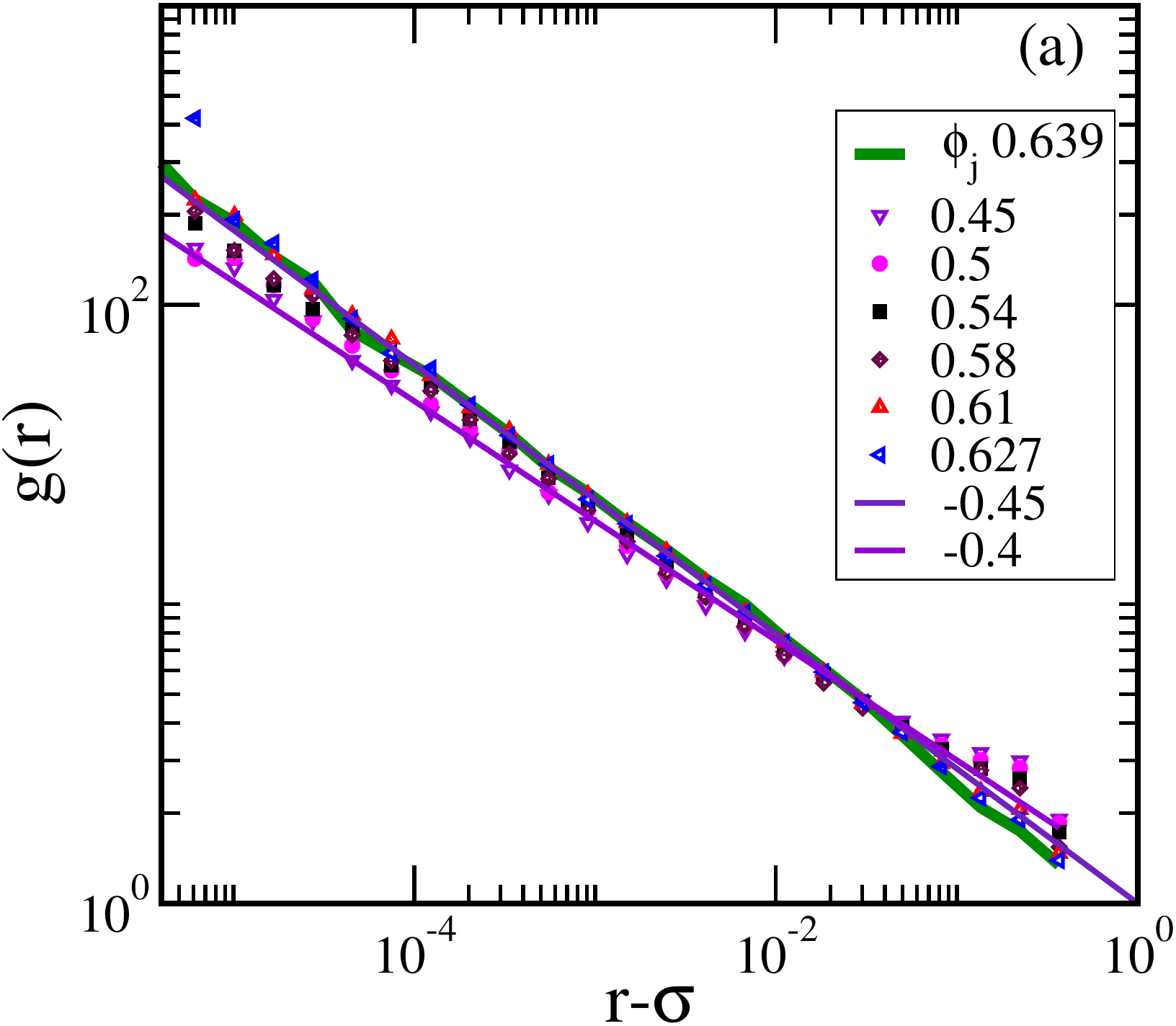}
\hskip 10mm
\includegraphics[width=7cm,height=7cm]{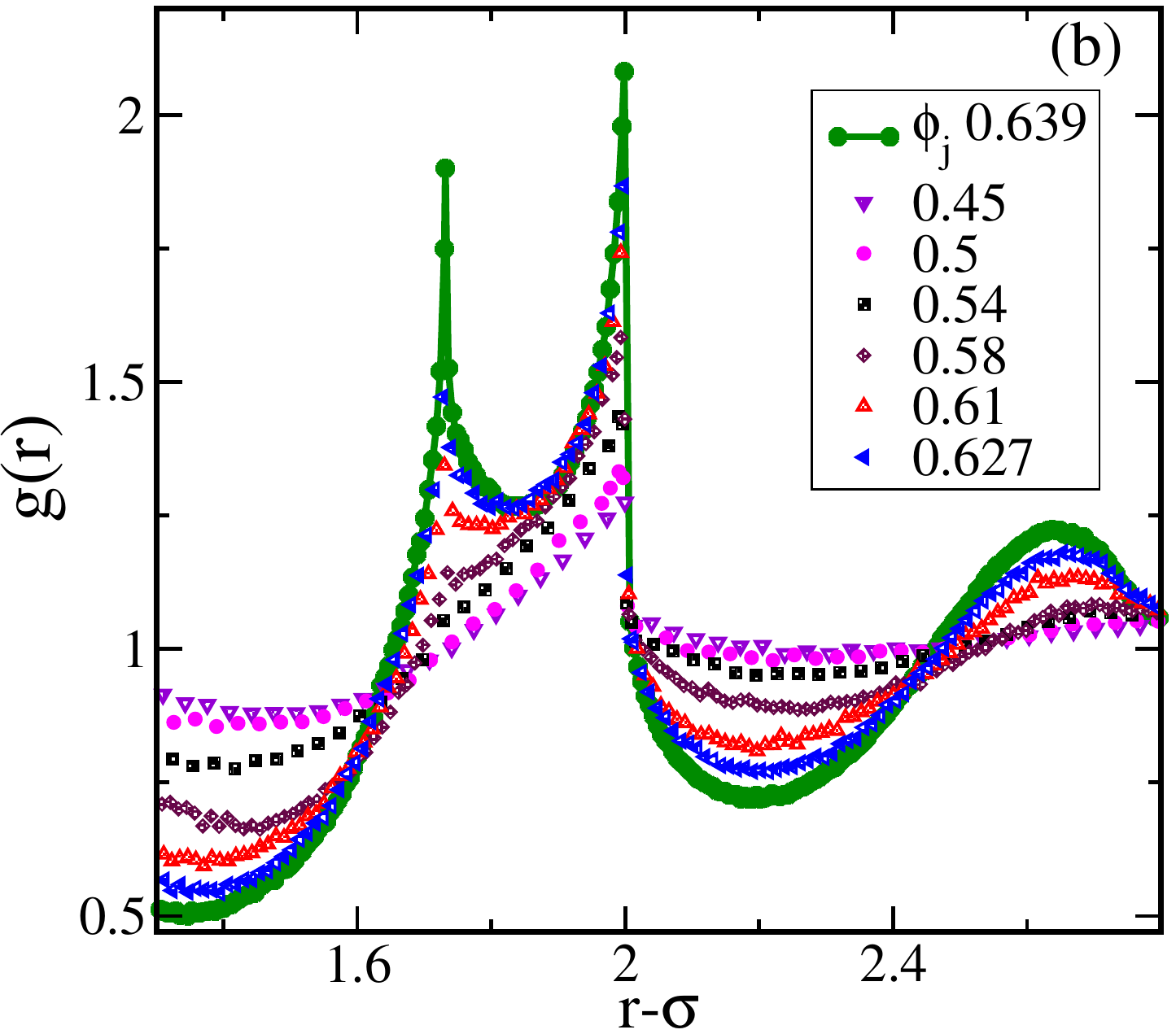}
\\
\vspace{1cm}
\includegraphics[width=7.4cm,height=7cm]{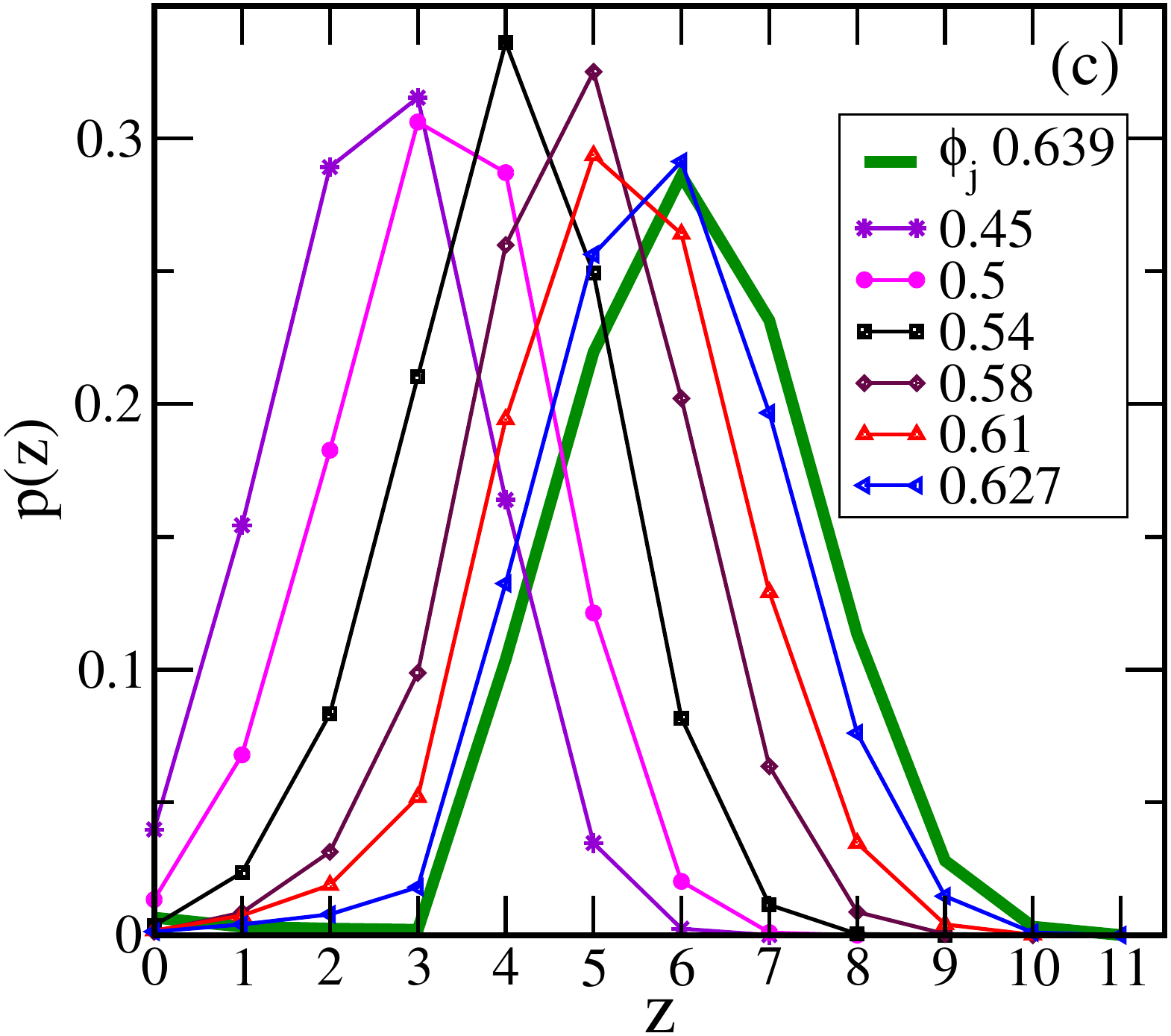}
\hskip 6mm
\includegraphics[width=7.4cm,height=7.1cm]{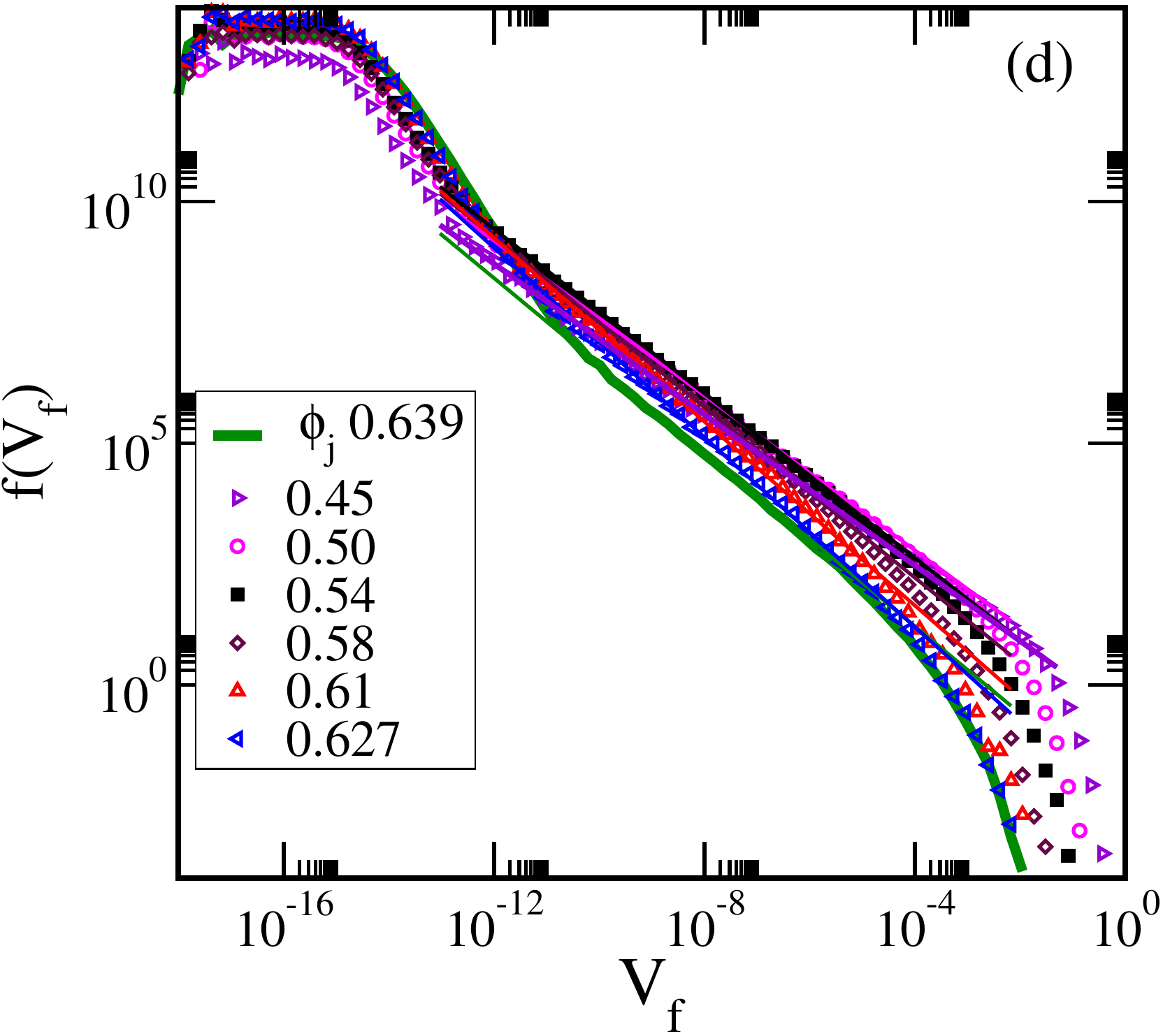}
\caption{\label{fig2} }
\end{figure}

We monitor the evolution of the structure under shear by considering
the pair correlation function $g(r)$, the distribution of the number
of contacts each sphere has, and the free volume distribution, each of
which exhibit unique signatures near the jamming point for
frictionless sphere packings: The pair correlation function exhibits a
near-contact power law singularity $g(r) \sim ({r\over \sigma} -
1)^{-\gamma}$ and singularities in the (split) second peak
\cite{ohern-2003,donev-2005}. The contact number distribution, \newtext{which we compute with and without considering rattlers (rattlers are spheres with fewer than $4$ contact neighbors)}, is
peaked at $Z = 6$, the value required by the isostaticity
condition. \newtext{The free volumes of individual spheres are computed using an exact 
algorithm that employs the Voronoi tessellation, and the free volume distribution exhibits a distinct power law tail  for nearly jammed packings, as described \cite{maiti-2014} and references therein}. This feature of the
free volume distribution has also been observed for sheared
configurations in two and three dimensions at high densities close to
the jamming point \cite{maiti-unpub}.  In Fig. \ref{fig1}, we show how
these features evolve, for sheared configurations at different values
of strain for packing fraction $\phi = 0.58$. It is seen that the
$g(r)$ develops a near contact power law \cite{ohern-2003,donev-2005},
initially absent, as shear strain is increased (Fig. \ref{fig1}(a)),
and the split-second peak develops the characteristic twin
singularities (Fig. \ref{fig1}(b)) \cite{donev-2005}. The peak of the
distribution of contact numbers, initially at zero, evolves to larger
values. Finally, the free volume distribution, initially exhibiting a
form typical of the fluid, develops a power law tail characteristic of
nearly jammed packings \cite{maiti-2014}. Thus, sheared fluid
configurations at $\phi = 0.58$ develop, by all these measures,
characteristics of jammed configurations. Fig. \ref{fig2} shows the
same quantities in the steady state, for the range of densities
studied, demonstrating the same behaviour at all densities. While the
near contact power law in the $g(r)$ is very similar in all cases, the
sub-peak at smaller $r$ in the second peak, as may be expected from
packing considerations, becomes sharper as density increases. The
average number of contacts moves from values less than $4$ towards $6$
as density increases. Table I summarises information on contact
numbers. The exponent in the free volume distribution changes slightly
with density, as does the exponent describing the near contact in
$g(r)$. Although these variations require explanation, the main point
is clear: with an increase of shear strain, initially fluid
configurations over a broad range of densities evolve structures that
bear strong resemblance to jammed configurations.

\begin{figure}[h]
\includegraphics[width=7cm,height=7cm]{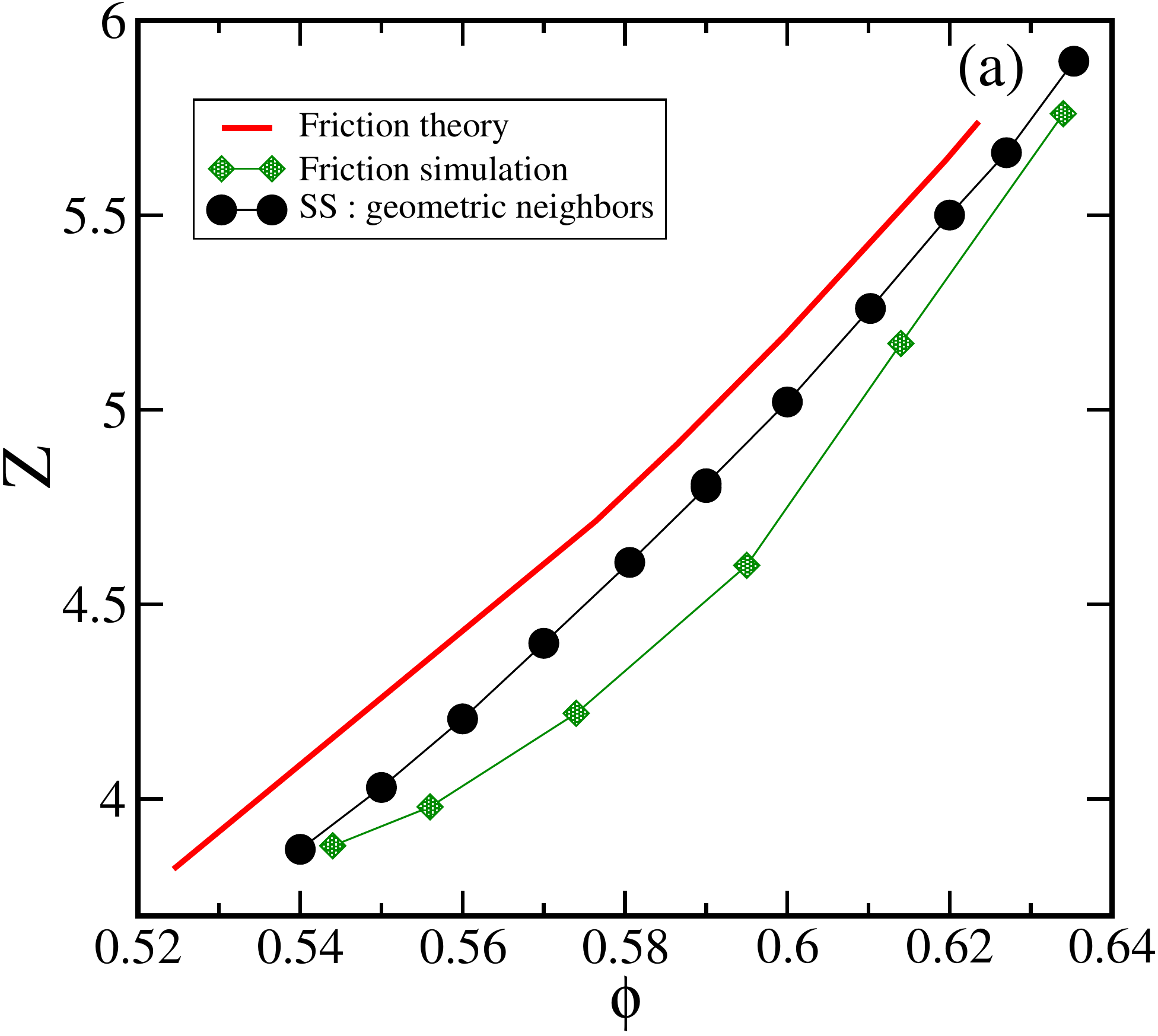}
\hskip 10mm
\includegraphics[width=7cm,height=7cm]{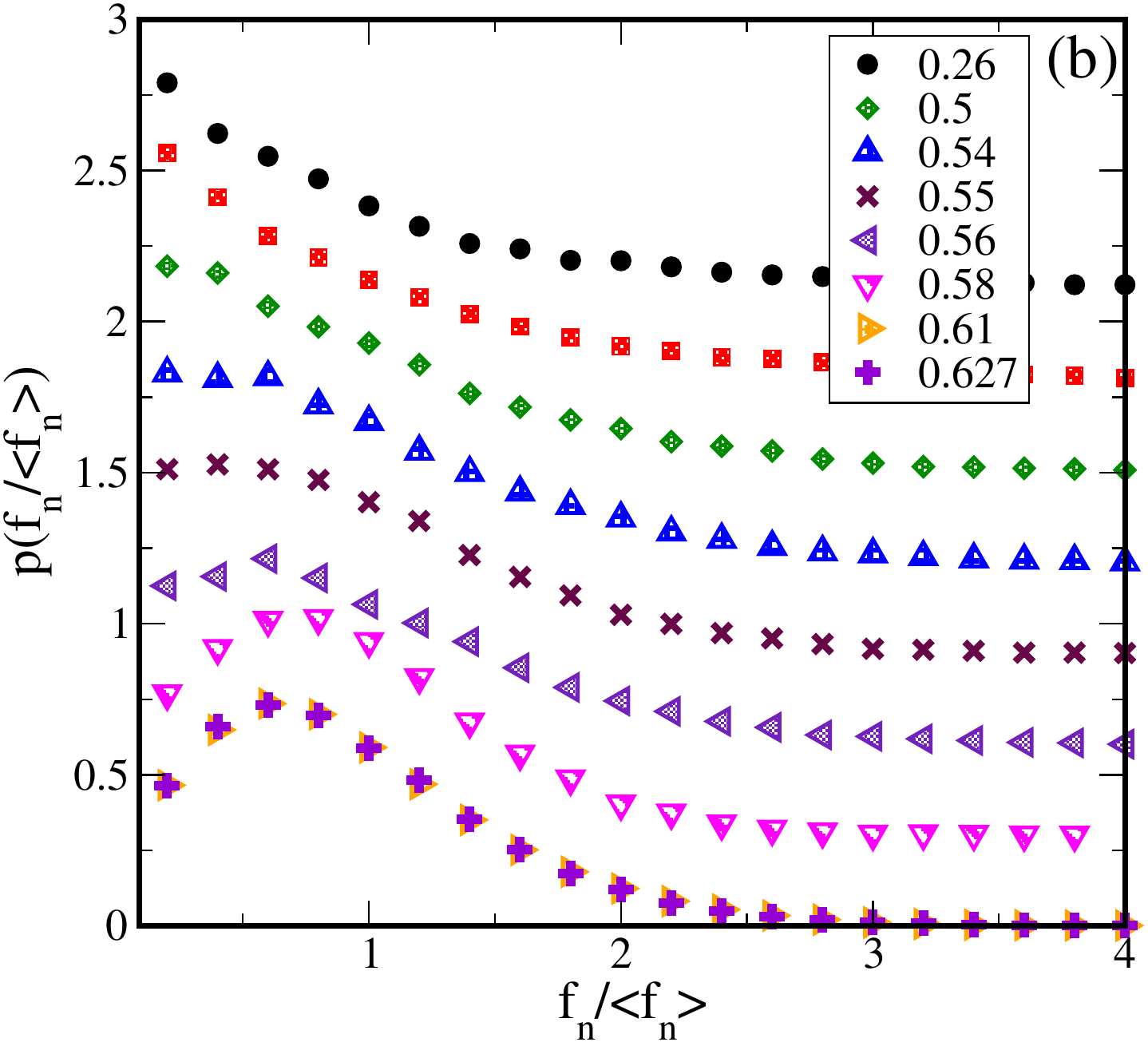}
\\
\vspace{1cm}
\includegraphics[width=7.2cm,height=7cm]{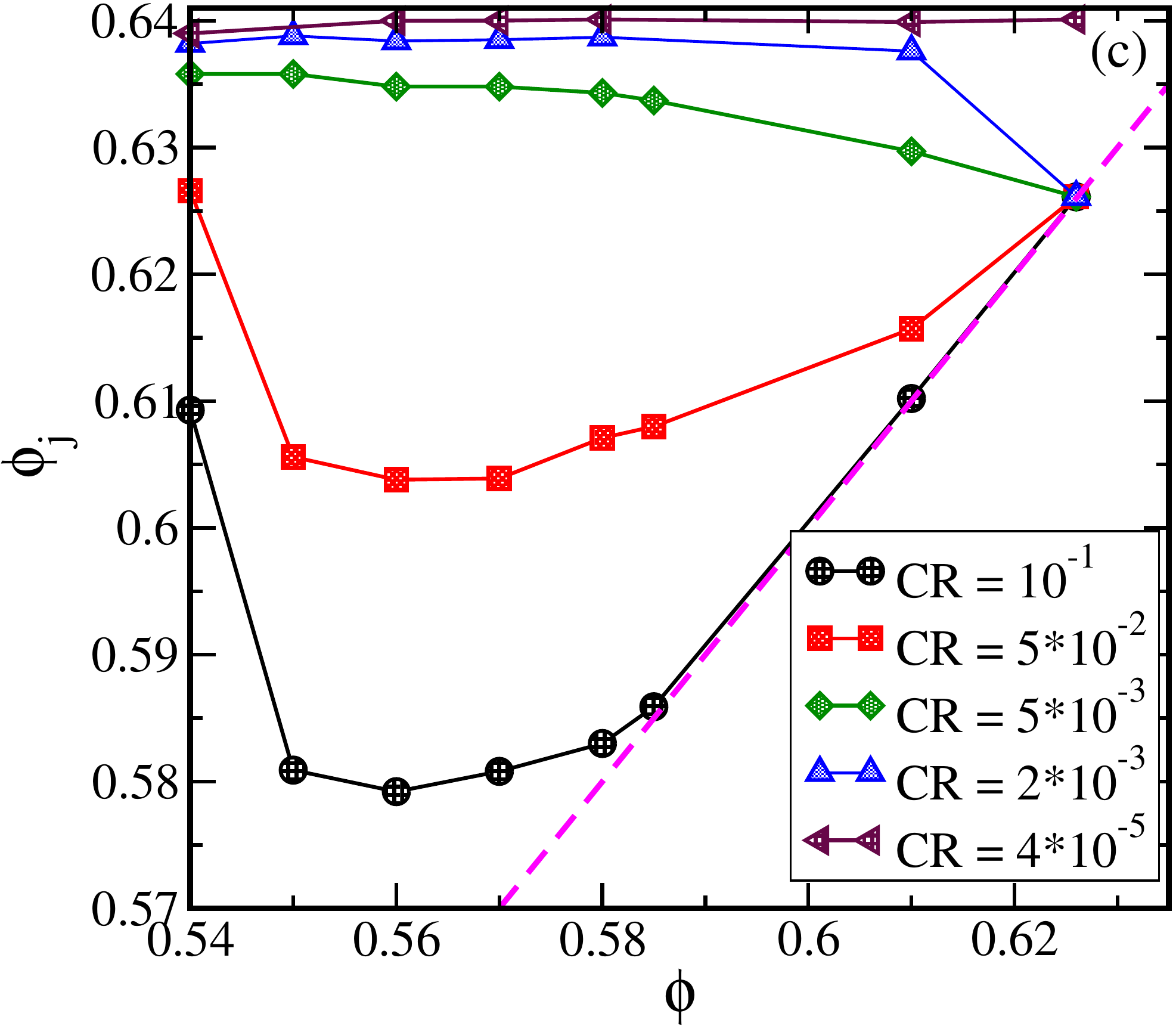}
\hskip 10mm
\includegraphics[width=7cm,height=7cm]{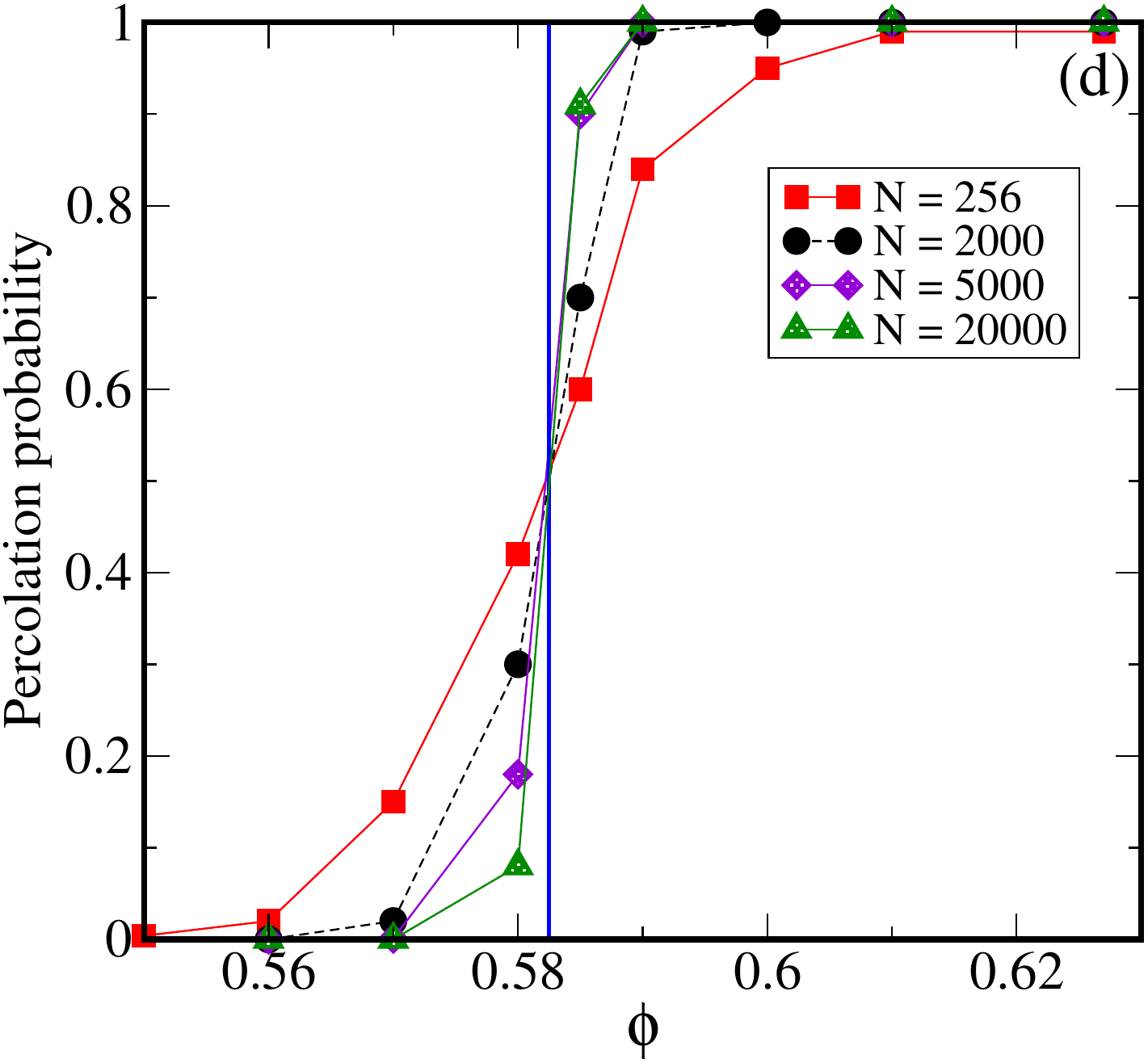}
\caption{\label{fig3} }
\end{figure}

We consider next the statistics of the number of contacts and the
distribution of forces between spheres. In Fig. \ref{fig3} (a), we
show the parametric relationship between the mean contact number and
the density that we have of our sheared configurations of frictionless
spheres. The mean contact number decreases from a $6$ at the isotropic
frictionless jamming density of $\sim .64$, to a value of $4$ at a
density $\sim 0.55$. Interestingly, the relationship displayed by our
data closely matches those of \cite{makse-2008,silbert-2010}, for
frictional jammed packings. The presence of a threshold density of
$0.55$ is further supported by the distribution of contact forces
normalised to their mean value, $P(f)$, shown in Fig. \ref{fig3}
(b). These distributions show a peak at finite values above $\phi =
0.55$, a previously identified characteristic of jammed packings
\cite{ohern-2001,wyart-2012,lerner-2013}. These results together
indicate the presence of a threshold density $0.55$, and a similarity
between the sheared configurations of frictionless spheres we generate
and frictional jammed packings. This observation has resonance with a
number of other past suggestions, including the occurrence of a glass
transition \cite{speedy-1998}, shear thickening \cite{brown-2009},
onset of dilatancy and random loose packing \cite{onoda-1990}. The
last of these, random loose packing, however, is a feature of
frictional packings, and we have so far dealt with frictionless
packings. An appealing picture, which we explore below, is that shear
deformation, even in the absence of friction, serves to induce
structures \cite{cates} at densities above $0.55$, \newtext{and considerably below $\phi_J$}, which may be stabilised by friction, when present, to produce shear jammed packings. 

Before we consider the role of friction, we consider the jamming
properties of the sheared frictionless packings themselves, by
applying the Lubachevsky-Stillinger \cite{LS} jamming procedure (see
Methods) for a range of compression rates to the configurations
generated at different densities. For slow compression rates, for all
initial densities, the resulting jammed configurations have densities
close to $\phi = 0.64$, as seen in Fig. \ref{fig3} (c). However, at
high compression rates, we note a difference in behaviour across $\phi
\sim 0.58$. At higher initial densities, configurations jam at roughly
the same density, while at lower initial densities, the jammed
densities are higher. The possibility of generating jammed
configurations above a density of $0.58$, albeit at high compression
rates, appears related to the percolation of locally stable
structures.  In Fig. \ref{fig3} (d) we show the percolation
probability of spheres with contact number $Z \ge 2 D = 6$, for
different system sizes $N = 256, 2000, 5000$. The connected clusters
of such spheres percolations at $\sim 0.58$. The suggested possibility
of $0.58$ being a threshold that is distinct from that at density
$0.55$, and the analysis of other percolation characteristics that may
elucidate the limiting density of $0.55$ suggested by other results
above, deserve further analysis but are not pursued further here. 
\newtext{Percolation of $D + 1$ coordinated spheres occurs at a much lower density, data for which is shown in the SI for completeness.} 

\begin{figure}[h]
\includegraphics[width=7cm,height=7cm]{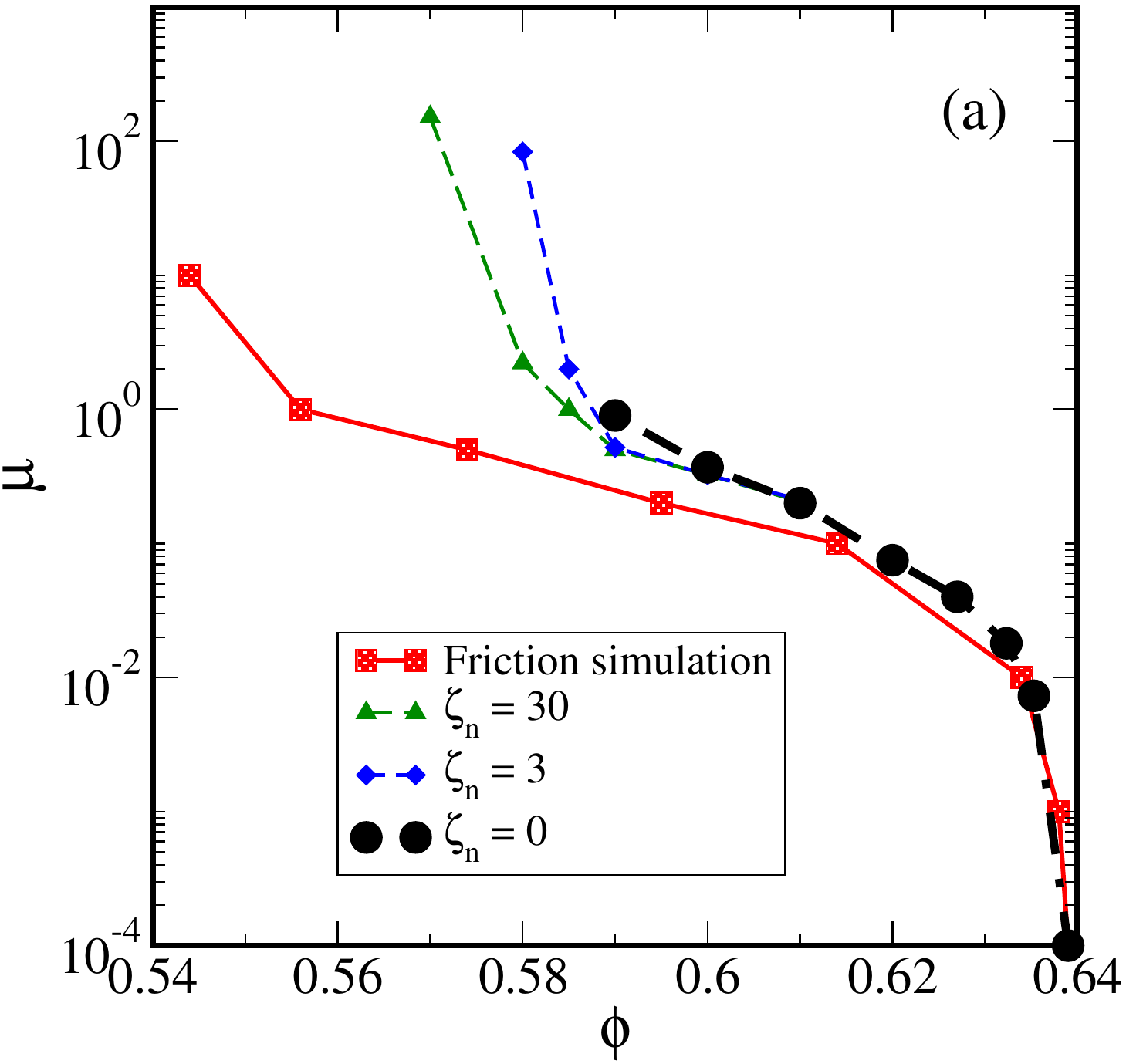}
\hskip 10mm
\includegraphics[width=7cm,height=7cm]{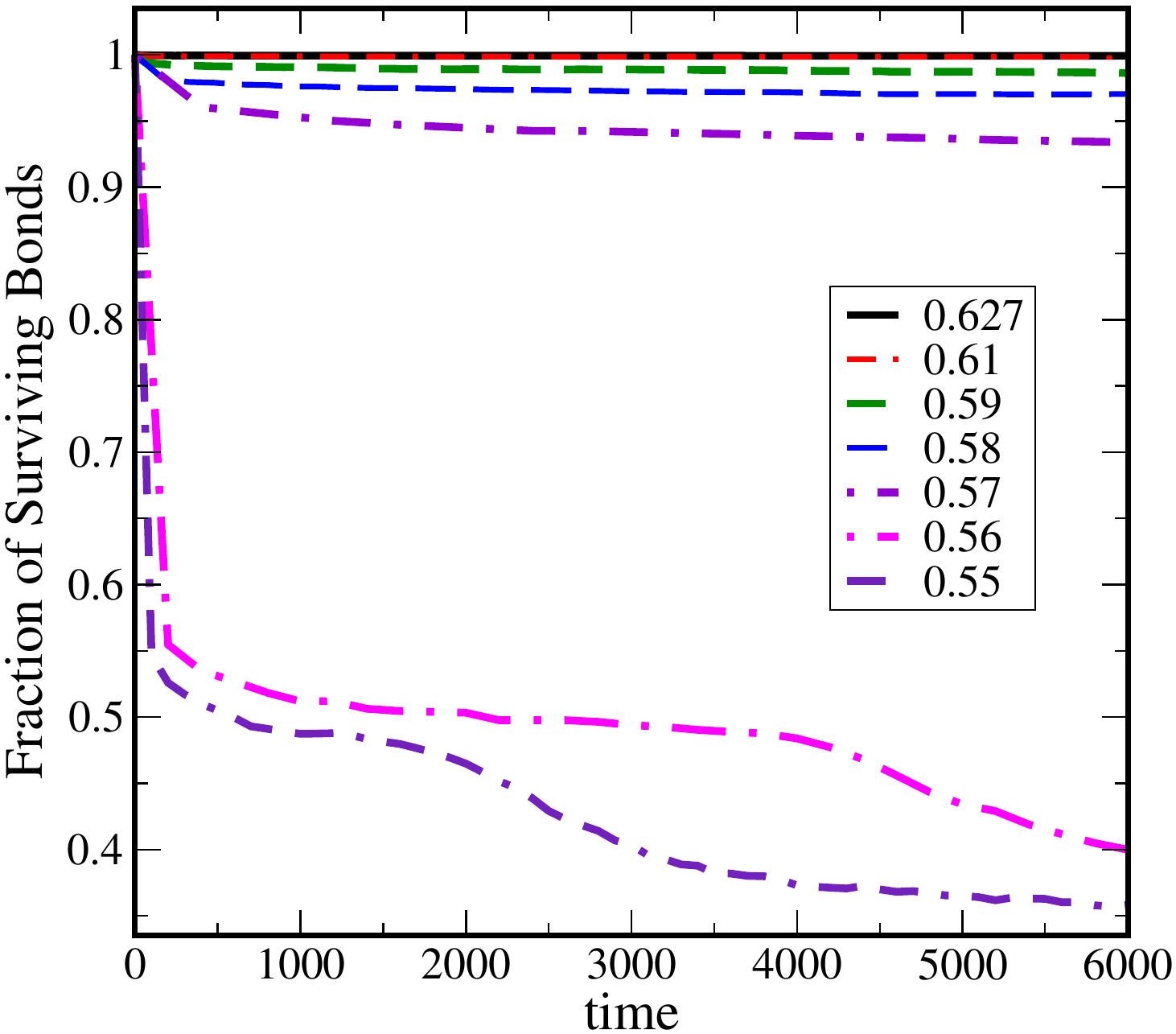}
\vspace{1cm}
\includegraphics[width=7.2cm,height=7cm]{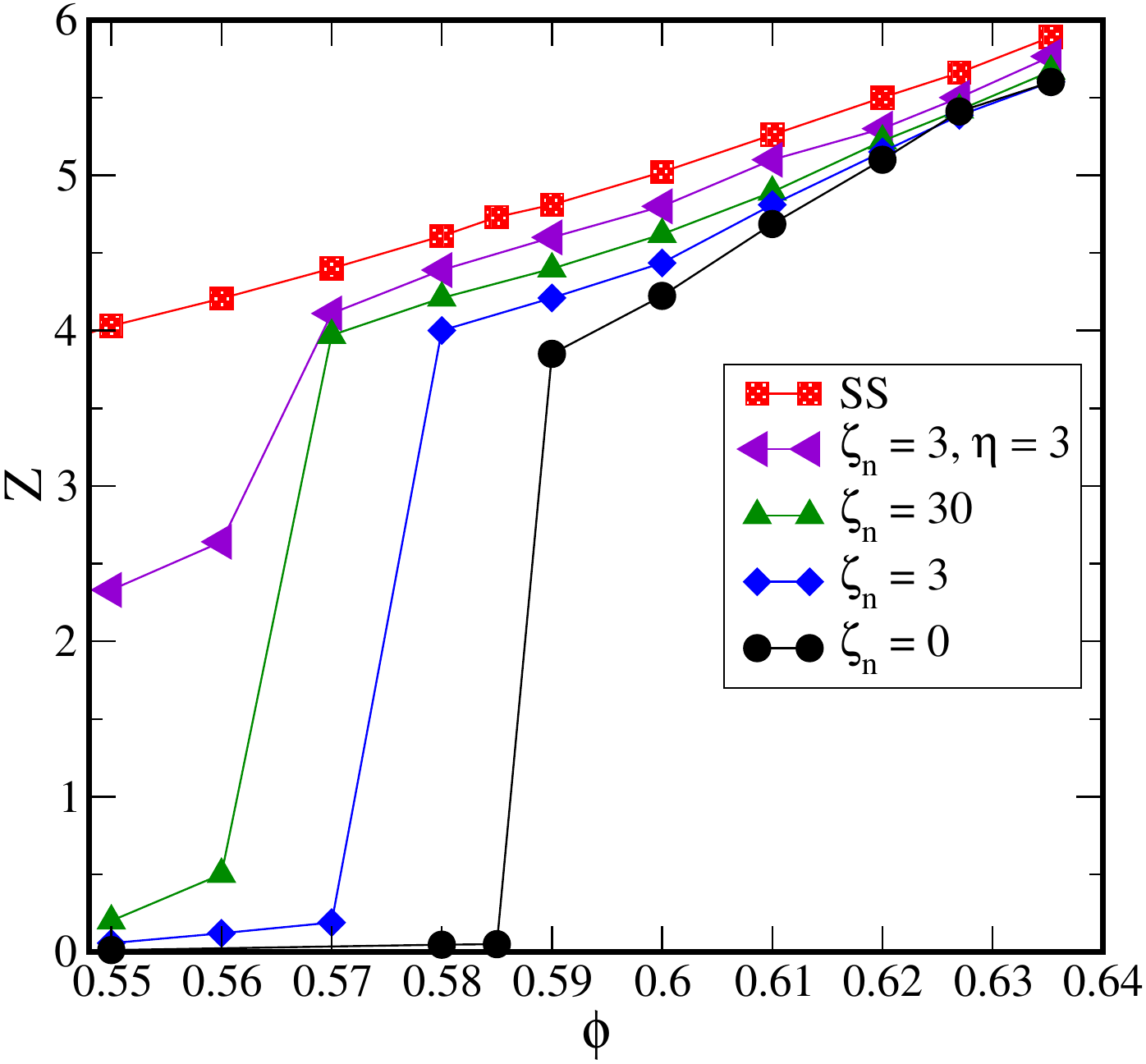}
\hskip 10mm
\includegraphics[width=7cm,height=7cm]{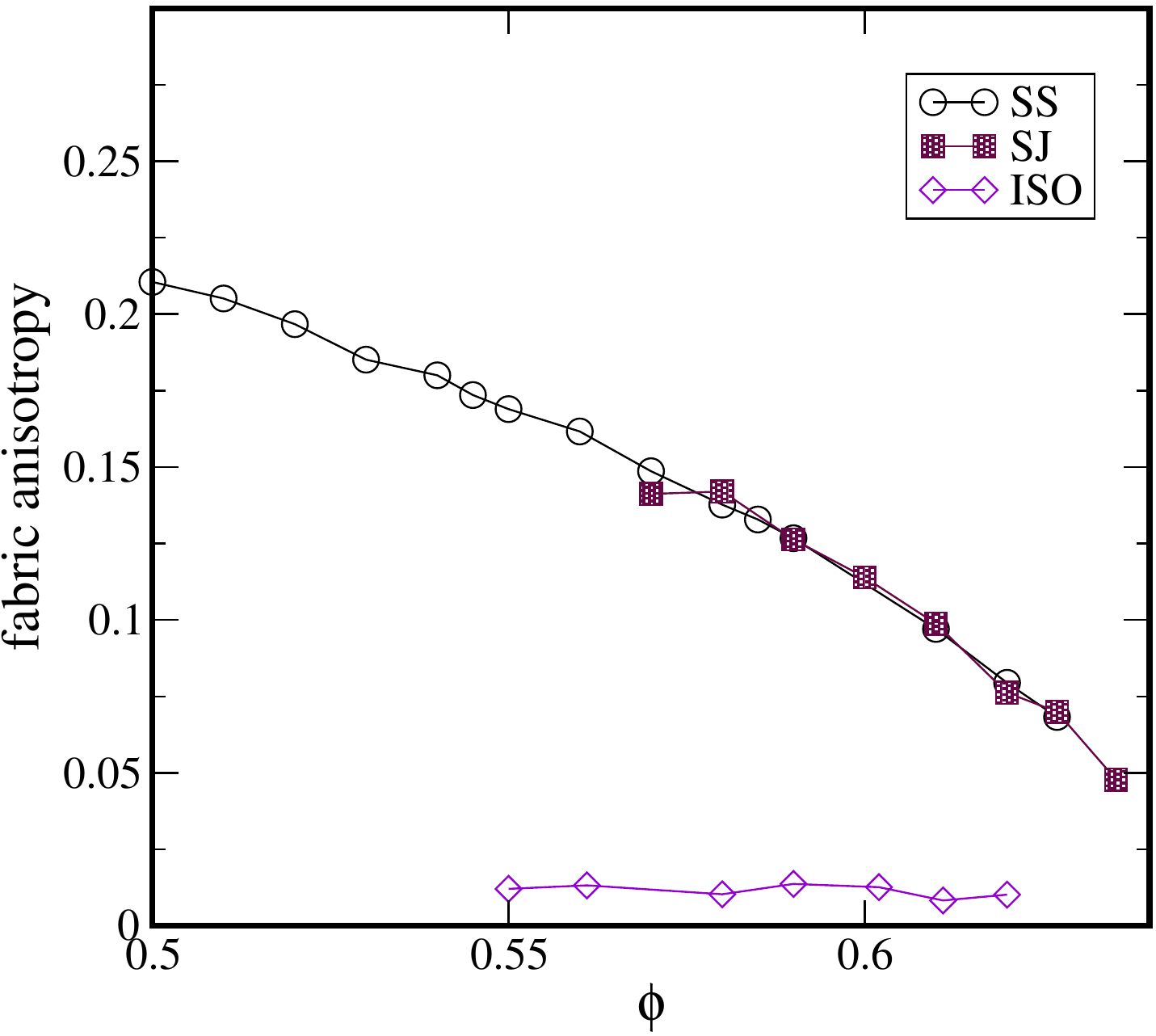}
\caption{\label{fig4}}
\end{figure}

In order to assess whether sheared frictionless spheres can jam in the
presence of friction, we perform simulations including frictional
contact forces using the discrete element method (DEM). We subject the
steady state sheared configurations to a strain step ($d\gamma =
5\times10^{-5}$) in addition to slight compression (see Methods and
SI), and evolve them using DEM, varying the friction coefficient $\mu$
and the damping coefficients $\zeta_n$ and $\zeta_t$ (see Methods
and SI), and monitor the evolution of the structure. We initially
choose damping coefficients $\zeta_n = \zeta_t = 0$. Although the
spheres do not move significantly during any of these simulations
(with mean squared displacements less than $10^{-4}$), for small
enough friction coefficients at any density, the shear stress and the
average contact number decay rapidly to zero indicating that the
structure is unjammed (see SI). The threshold friction coefficient is
identified at each density beyond which the sheared configurations
remain jammed. For densities above $\phi = 0.58$, sheared
configurations remain jammed, while below, the configurations unjam
for the studied range of friction coefficients. Fig. \ref{fig4}(a)
shows the threshold friction coefficients obtained, which compare
reasonably well with values obtained in frictional simulations by
Silbert \cite{silbert-2010}. Fig. \ref{fig4} (b) shows the fraction of
initial contacts that survive as a function of time,  \newtext{for large friction coefficients well above the threshold value, with steady state configurations generated with $d\gamma = 10^{-12}$.} It is seen that most of the contacts remain
intact for densities above $\phi = 0.57$, while they decay to $0$
below.  A closer agreement with frictional simulations by Silbert
\cite{silbert-2010} are obtained by the inclusion of damping (as done
in \cite{silbert-2010}), as shown in Fig. \ref{fig4}(a). For $\zeta_n
= 3$ and $30$ ($\zeta_t = \frac{1}{2} \zeta_n$) respectively,
frictional jamming occurs down to $\phi = 0.58$ and $0.57$.   The mean
contact number $Z$, at the lower density limit to frictional jamming
reaches $4$ in all these cases \newtext{as shown in Figure \ref{fig4}(c).}
With a suitable procedure, we therefore expect the lower density limit
to frictional jamming to be $\phi = 0.55$, at which we obtain sheared
frictionless packings with $Z = 4$.  Thus, the sheared configurations
of frictionless spheres we generate jam in the presence of friction
above a threshold friction coefficient that closely matches those of
frictional packings \cite{silbert-2010}, above the threshold density
of $0.55$. \newtext{Shear jammed configurations form from sheared
  steady state configurations with negligible rearrangement, in
  contrast to isotropic frictional jamming, and are anisotropic, as
  shown in Figure \ref{fig4}(d). The anisotropy (defined from the
  fabric tensor; see Methods) of the shear jammed configurations is
  identical to the sheared steady state configurations in the range of
  densities where we obtain shear jamming.} \newtext{As an independent test, to be described elsewhere, we solve the force balance equations for the steady state configurations using the DEM generated forces as initial guesses, to obtain forces needed for force balance from geometric information alone.}

Although many questions are suggested from the results above that must
be investigated further, they show that shear deformation of spheres
even in the absence of friction at densities well below the isotropic
jamming point leads to the emergence of geometric features resembling
jammed packings, with a threshold density that may be identified with
the random loose packing limit that may be identified with the case of
infinite friction. The force distributions and the relationship
between the packing fraction and contact number in the steady state
support the comparison of the sheared configurations above the
threshold density with frictional jammed packings with varying
friction. Our results thus serve to disentangle the role of structure
formation under shear and friction in the generation of shear jamming
phenomenology. They also identify the lower limit of shear jamming
with random loose packing.
Whether the
same kind of structure formation can arise in isotropic
compression\cite{shen-2012} rather than shear, analysis of the
anisotropies in sheared structures, and the role of finite shear rates
present some obvious questions to pursue in future work.

\noindent{\bf Methods:}

The model system we study is composed of $N=2000$ frictionless spheres
interacting with soft harmonic repulsive potential, $v(r) = {\epsilon
  \over 2} \left( 1 - (r/\sigma)\right)^2$, where $\epsilon$ and
$\sigma$, the interaction strength and size of the spheres, define the
reduced units used throughout. The initial configurations are hard
sphere configurations, obtained at high densities in two ways: (1)
Starting from an equilibrated hard sphere fluid at initial density
$0.45$, a fast initial compression is effected using a Monte Carlo
simulation till the desired density is reached for the initial
configurations. (2) Starting from packings at $\phi_j$, obtained by
the Lubachevsky-Stillinger (LS) jamming protocol \cite{LS}, lower
density configurations are obtained by rescaling the simulation box
size.  The LS procedure involves event driven molecular dynamics of
hard spheres, whose radii are inflated at a specified rate. The
procedure terminates when the radii cannot be increased by any finite
amount without the next collision of a pair of spheres intervening, or
when the collision rate diverges. In practice, the procedure is terminated when the sphere radii do not change by more than $10^{-10}$ between successive collisions. 

Athermal quastisatic simulations (AQS) are performed using LAMMPS
\cite{lammps}, which involves the following steps: (1) Affine
transformation of coordinates by a small step with $d\gamma =
5\times10^{-5}$. (2) Energy minimisation using the conjugate-gradient
method, employing Lees-Edwards periodic boundary conditions. This
procedure is used till steady states are reached. We also use a strain
step of $d\gamma = 5\times10^{-12}$ to shear steady state configurations
further to validate our contact definition (see below), to evaluate
contact forces and to perform frictional simulations. 

Data shown are averaged over $50-70$ initial independent
configurations, except Figure 4  which are averaged over $10$
configurations, Figure 3 (b) which is averaged for $20$ configurations and Figure 3 (d) which is averaged over $1000$
configurations.

We use a cut-off of $\sim 10^{-5}$, which is the distance at which
$g(r)$ deviates from the power law, to define {\it contact} neighbours
in order to compute the contact number Z. This cut-off is a precision
limit that is dependent on the strain step used in AQS, as we show in
SI by considering various $d\gamma$ down to $5\times10^{-12}$. In order to have a
consistent definition of contact, we compress configurations by
rescaling the diameter of the spheres (by $\sim 10^{-5}$ when $d\gamma
= 5\times10^{-5}$ and $\sim 10^{-12}$ when $d\gamma = 5\times10^{-12}$ {\it
  etc.})  so that all neighbor pairs identified as contact neighbors
have finite contact forces. Since this equivalence becomes more exact
for smaller strain steps, we consider steady state configurations with
strain step $d\gamma = 5 \times 10^{-12}$ when evaluating forces between
contact neighbours (Fig. 3(b)). The data for frictional simulations (Fig. 4) is for $d\gamma = 5\times10^{-5}$. \newtext{Data shown in the SI validate in detail the procedure we adopt.} 

In order to generate jammed configurations using the LS procedure, the
small overlaps in the sheared configurations are removed by decreasing
the diameter by a small amount ($\sim 10^{-9}$) which is then
increased to unity through the LS protocol with a fast compression
rate of $0.1$, involving negligible displacements of spheres ($\sim
10^{-16}$). We use the steady state configurations to calculate free
volumes of the particles using the algorithm described in \cite{maiti-2014}. 

To test the stability of steady state sheared structures in presence
of friction, we use discrete element method (DEM) \cite{cundall} to
model contact interactions between particles through a repulsive
linear spring-dasphpot model. The model, described further in SI,
involves normal and tangential spring constants $\kappa_n$,
$\kappa_t$, damping coefficients $\zeta_n$, $\zeta_t = \frac{1}{2} \zeta_n$, and the
friction coefficient $\mu$ as parameters.  The model parameters used
are $\kappa_n$ = $\kappa_t$ = $2$ and the normal contact damping $\zeta_n = 0, 3, 30$.
At each contact, the Coulomb yield criterion is obeyed i.e,
$F_t \le \mu F_n$, where $\mu$ is the friction coefficient which is
initially varied from $.01$ to $100$ in multiple of $10$ in order to
bracket the threshold value beyond which configurations are
jammed. Threshold $\mu$ values are refined further by considering a
finer grid of values. Configurations identified as jammed display a
finite shear modulus, which we illustrate in one case in the SI. 

\newtext{Anisotropy of initial and jammed structures for isotropic and sheared steady state initial conditions is calculated, using the fabric tensor, defined as
\begin{equation}
 \hat{R} = \frac{1}{N}\Sigma_{i \neq j} \frac{{\bf r}_{ij}}{\mid r_{ij}\mid} \otimes \frac {{\bf r}_{ij}}{\mid r_{ij}\mid}
\end{equation}
where ${\bf r}_{ij}$ are distance vectors between contact
neighbors. The normalized difference between the largest eigen value
$C_{1}$ to the smallest $C_{3}$, $(C_{1} - C_{3})/(C_{1}+C_{2}+C_{3})$ defines the fabric anisotropy.}

%


\bigskip

We wish to thank S. S. Ashwin, Bulbul Chakraborty, Corey O'Hern, Stefan Luding, Itamar Procaccia and Sidney Nagel for useful discussions. We gratefully acknowledge the use of computational resources at TUE-CMS, JNCASR, Bangalore and TCIS, TIFR Hyderabad. This research was supported in part by the National Science Foundation under Grant No. NSF PHY11-25915. 

\eject 

\centerline{\LARGE \bf Figure Captions} 

\noindent{\bf Figure 1 : (a)} Evolution
of radial distribution function $g(r)$ for $\phi = 0.58$ with strain
($\gamma$). For {\bf $\gamma = 0.41$}, the system has reached the
steady state and has a power law in $g(r)$ extending over $5$ decades.
{\bf (b)} Radial distribution function showing changes in the second
peak as the system is strained. As we move from the bottom curve to
the top curve, the direction of increase in strain, the discontinuity
in the second peak develops and becomes stronger. Each curve are
shifted from by $1$ from the previous along the $y$ axis for clarity.
{\bf (c)} Distribution of contact number as a function of
strain. 
\newtext{At zero strain,} there are no contact neighbours and after $\gamma =
0.1$, particles begin to have contact neighbours.  {\bf (d)} Evolution
of the free volume distribution with strain. The power law tail in
$f(V_f)$ develops around the same strain value for which contact
neighbour distribution develops a peak at a non zero value of $Z$.
Data for nearly jammed configurations at $0.639$ are shown in all
panels for comparison (thick green lines).  \\\\ 
\noindent{\bf Figure 2 : (a)
}Power law divergence in $g(r)$ for sheared configurations in the
steady state. The power law exponent value depends slightly on density
and is about $0.45$ for the nearly jammed case.  {\bf (b) }
Singularities in the second peak of the $g(r)$ are observed at all
densities at $\sqrt{3} \sigma$ and $2 \sigma$, with the feature at
$\sqrt{3} \sigma$ becoming stronger as density increases.  {\bf (c) }
Distribution of contact number.  {\bf (d)} Free volume
distributions. The exponent of the power law tail \newtext{indicated by fit lines} depends on slightly
on density. 
Data for nearly jammed configurations at $0.639$ are shown in all panels for comparison  (thick green lines).
\\\\ \noindent{\bf Figure 3 : (a) } Parametric plot showing the average
contact number {\it vs.} density for sheared configurations, compared
with that of frictional packings obtained by Song et al
\cite{makse-2008} and Silbert \cite{silbert-2010}.
 The average contact number becomes greater than $4$ for $\phi \ge 0.55$. 
{\bf (b)} The distribution of contact forces for different
densities which show a peak at finite values for $\phi \ge 0.55$. 
 {\bf(c) } Configurations subjected to the LS jamming protocol at
different compression rates. Plotted are the resulting densities of
jammed configurations indicating a change of behaviour for $\phi > 0.58$. Compression rates in the legend are indicated by "CR". 
 {\bf (d)} Percolation of jammed ($Z
\ge 2D$) spheres in the sheared configurations indicating a
percolation threshold at $\phi = 0.585$, shown for different system
sizes. \\\\
\noindent {\bf Figure 4 : (a)} Threshold friction coefficient as a function of the density, beyond which sheared steady state structures jam, for damping coefficients $\zeta_n = 0, 3, 30$. Also shown for comparison are results for frictional packings obtained by Silbert \cite{silbert-2010}. 
 {\bf (b) } Fraction of contact pairs initially present that survive as a function of time, for the threshold friction coefficient (for density $\phi \geq 0.59$; for $\phi = 0.55$ and $0.58$ the friction coefficient is $10$), and $\zeta_n = 0$. \newtext{{\bf (c) } Comparison of coordination number $Z$ of the  initial structure and the final structure obtained from frictional dynamics after applying a strain of $\Delta \gamma =  5 \times 10^{-5}$. Curves labeled $\zeta_n = 0, 3, 30$ are for steady state configurations with $d\gamma = 10^{-5}$, just above the threshold friction coefficient, and no solvent friction, whereas configurations for the curve labeled $\zeta_n = 3, \eta = 3$ are obtained with $d\gamma = 10^{-12}$, for friction coefficients well above threshold, and finite solvent friction $\eta$. The contact number if close to, but different from, the initial value for densities above $\phi = 0.57$, whereas they are zero for lower densities (except the case $\zeta_n = 3, \eta = 3$, where overdamping results in cessation of paricle motion before all contacts are lost). {\bf (d) } Fabric anisotropy as a function of packing density for steady state (SS), shear jammed (SJ) and isotropic frictional jammed (ISO) configurations. The anisotropy of SS and SJ packings are same.}

\eject

\begin{table}[t]
\caption {Statistics of steady state structural features of sheared
  packings. $\phi$ is the density of sheared packings, $<Z>$ and
  $<Z_{NR}>$ is the average coordination number with and without
  rattlers, and RP is the rattler ($Z \le 3$) percentage. $Q_6$ is the global bond orientational parameter computed to check absence of crystallinity. \\}

\begin{tabular}{|l|l|l|l|l|l|l|l|l|}
\hline
\hline
$\phi$ &  0.45 &   0.5  &  0.54   &  0.56 &   0.58  &  0.59  &  0.61  &  0.627 \\ [1ex]
 \hline
$<Z>$  &  2.54 &   3.23 &   3.88 &   4.24   & 4.615 &   4.81  &  5.26 &   5.65 \\ [1ex]
 \hline
$Q_6$  &  0.049  & 0.054  & 0.052  & 0.038  & 0.028  & 0.027  & 0.025  & 0.023 \\ [1ex]
 \hline
RP    &  0.795  & 0.572 &  0.358 &   0.262 &  0.175 &  0.138  & 0.078 &  0.037 \\ [1ex]
 \hline
$<Z_{NR}>$  &   4.2   &  4.38  &  4.63 &   4.82  &  5.042 &  5.175  & 5.495  & 5.785 \\ [1ex]
\hline
\hline
\end{tabular}
\end{table}

\clearpage

\centerline{\LARGE \bf Figures}

\pagebreak

\pagebreak

\pagebreak

\pagebreak 

\begin{center}
\textbf{Disentangling the role of structure and friction in shear jamming (Supplementary Information)}
\\
\textbf{H.A. Vinutha and Srikanth Sastry}
\end{center}


Here we present additional information regarding three aspects of our analysis of sheared sphere configurations, namely: (i) Approach to the steady state, (ii) Evaluation of the contact number, and (iii) Determination of the lower limit of shear jamming in the presence of friction. 


\section{Approach to the steady state}
\begin{figure}[h]
\includegraphics[scale=0.4,angle=0]{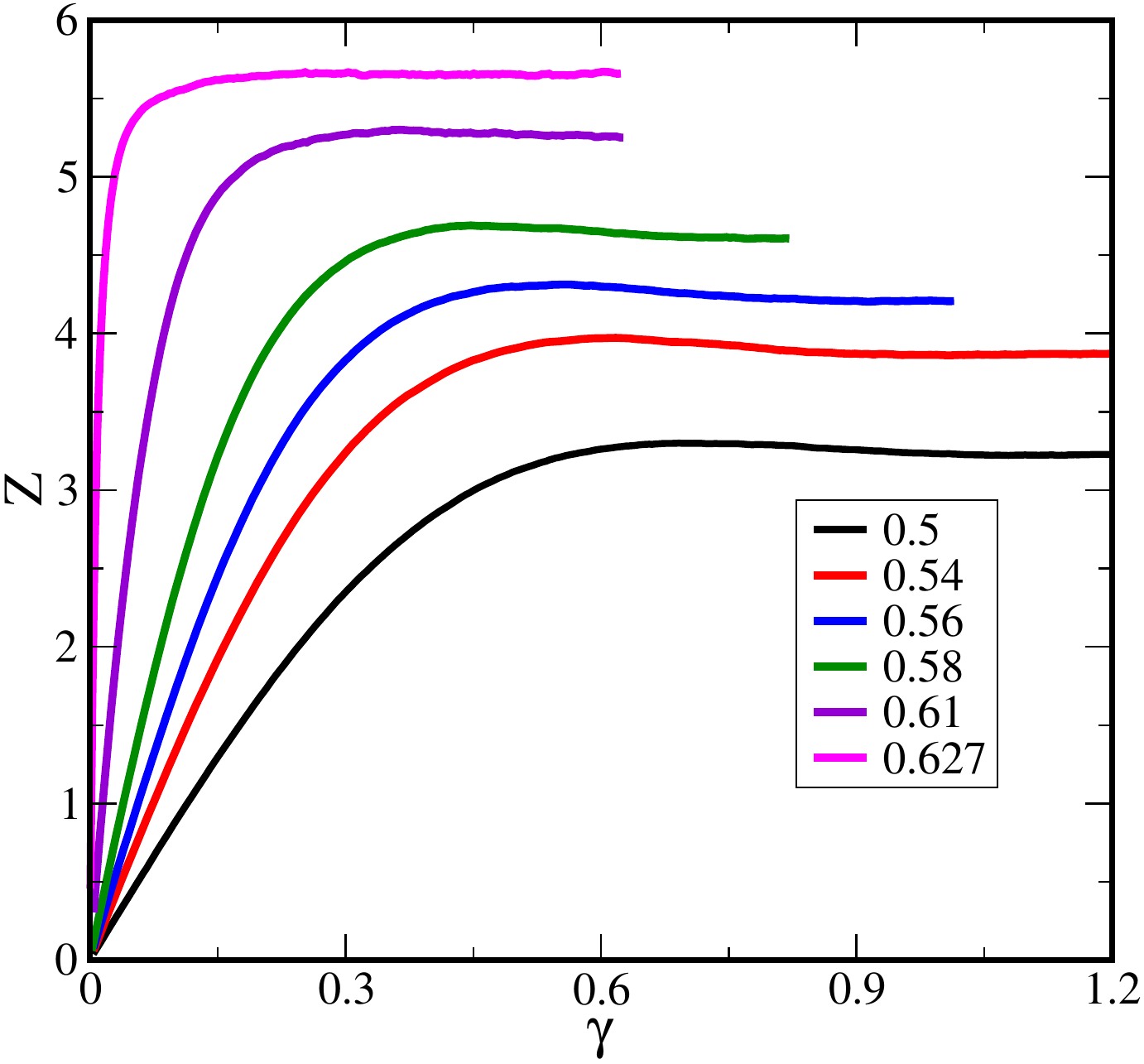}
\hspace{1cm}
\includegraphics[scale=0.4,angle=0]{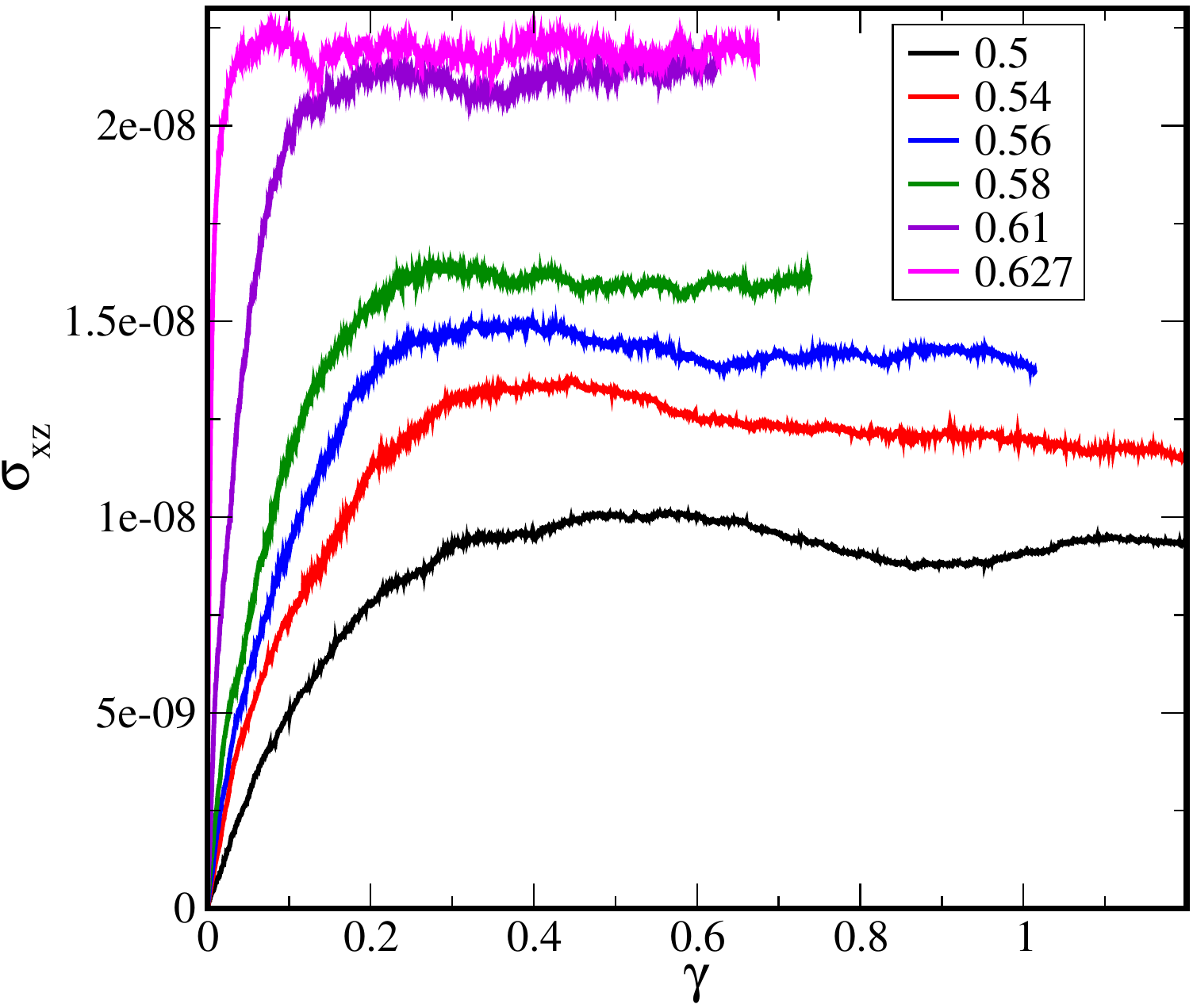}
\caption{\label{figSI1} {\bf (a)} Coordination number $Z$ as a function of strain for different densities. The value of $Z$ and the strain value where the steady state is reached depends on density. {\bf (b)} Stress in the shear plane as a function of strain for different densities.  The strain values at which steady state stress is reached compares well with the corresponding values for the coordination number $Z$.}
\end{figure}

The analysis we present is performed for steady state configurations. Whether the simulated system has reached steady state or not is assessed by monitoring various structural and dynamic quantities, which show behavior consistent with the average contact number and the shear stress. Because the stresses are negligible for energy minimum configurations, we calculate stresses for configurations before minimization in the AQS protocol. The  average contact number (whose evaluation is discussed next) and shear stress reach steady state values beyond strain values that depend on the density (Fig. \ref{figSI1}).

\section{Evaluation of the contact number $Z$}

\begin{figure}[h!]
\includegraphics[scale=0.4,angle=0]{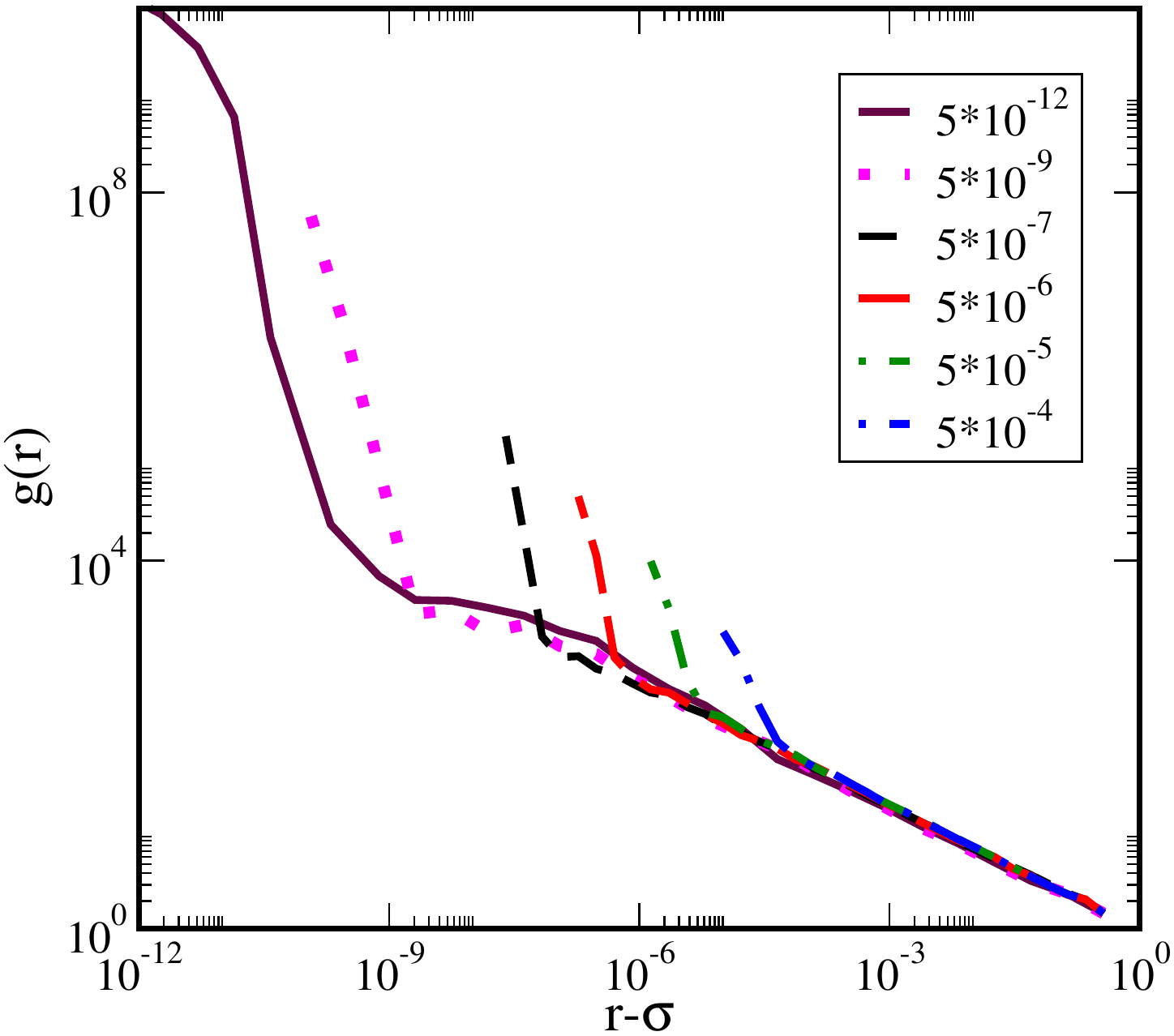}
\hspace{1cm}
\includegraphics[scale=0.4,angle=0]{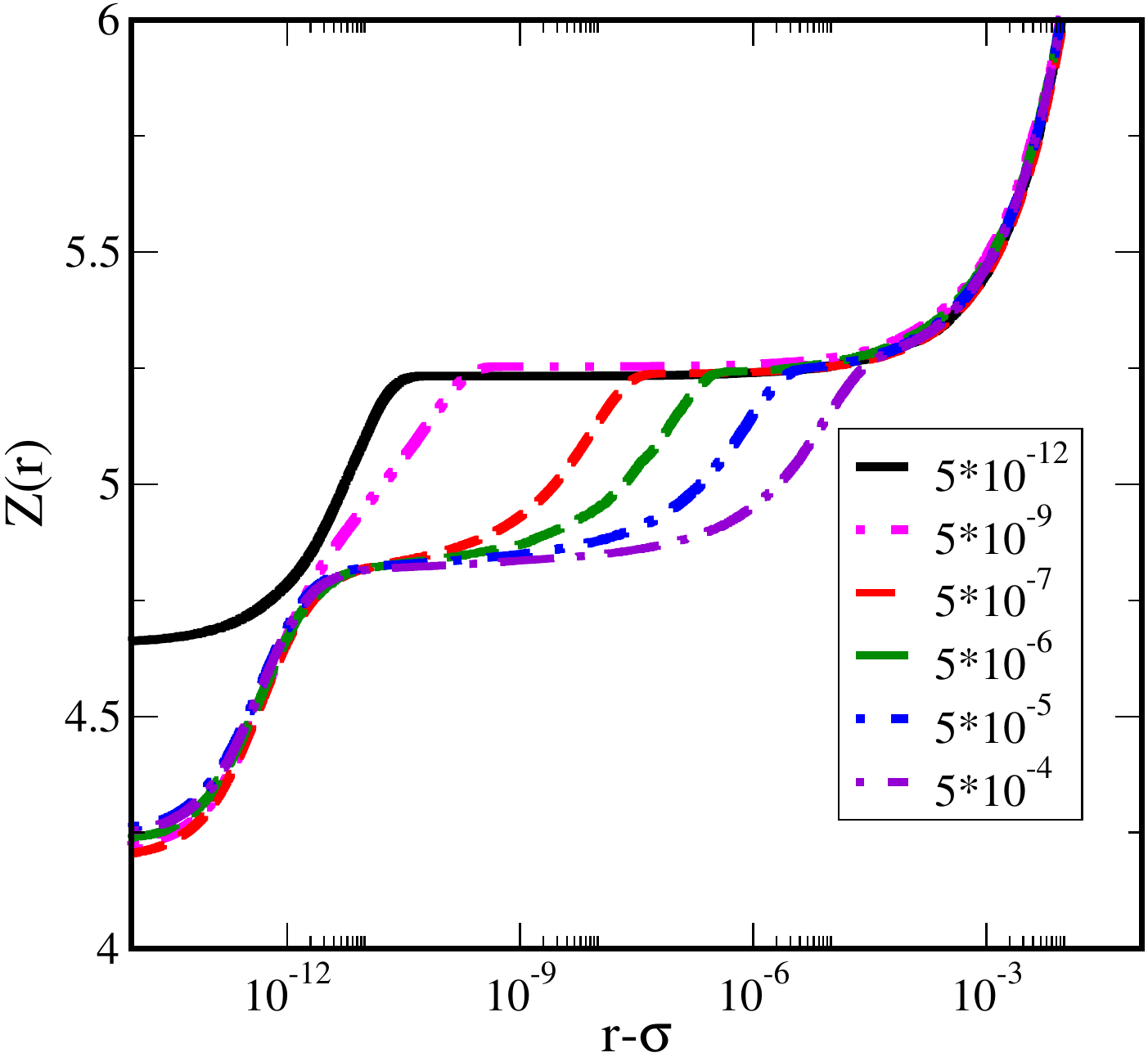}
\hspace{1cm}
\includegraphics[scale=0.4,angle=0]{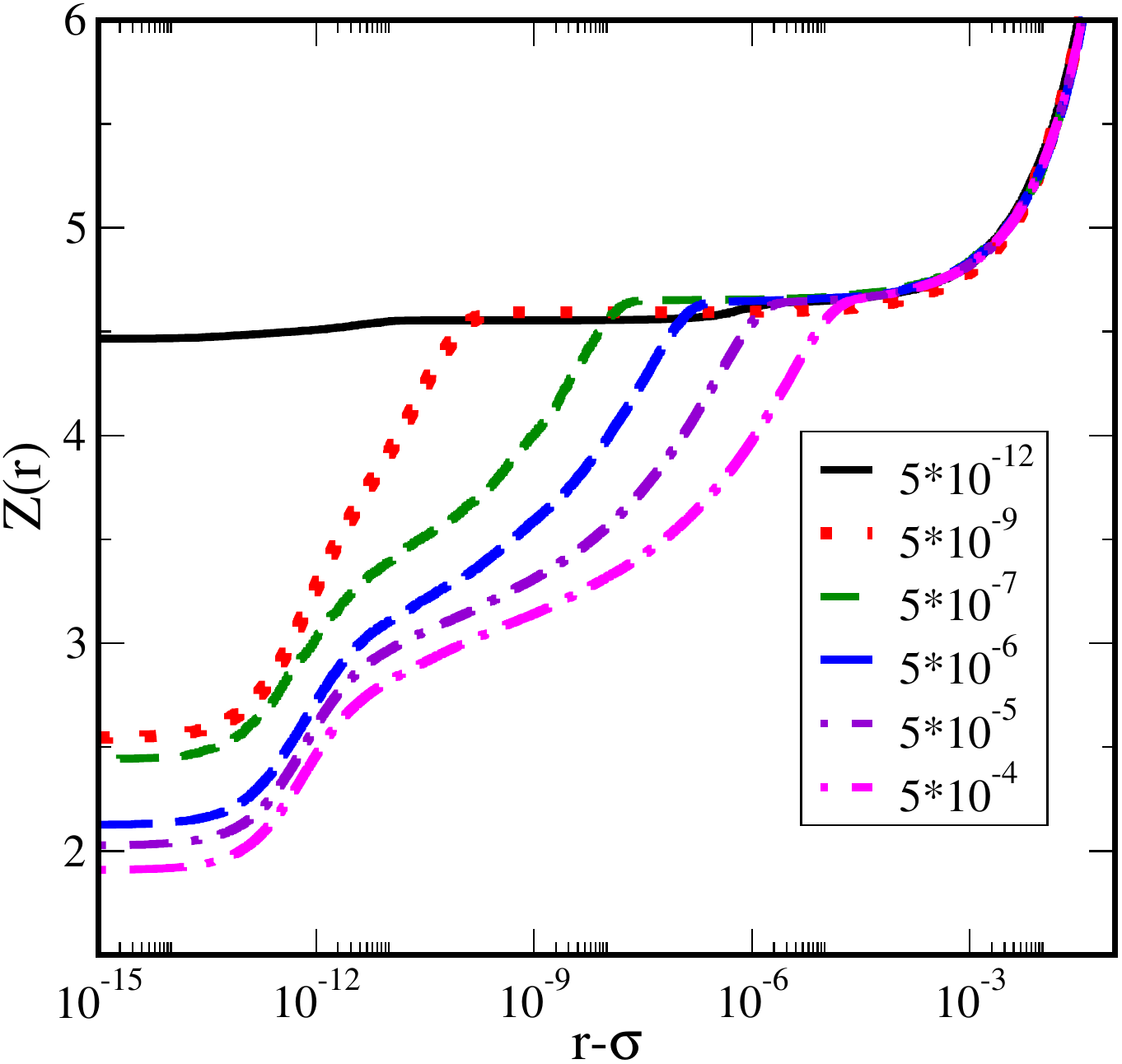}
\hspace{1cm}
\includegraphics[scale=0.3,angle=0]{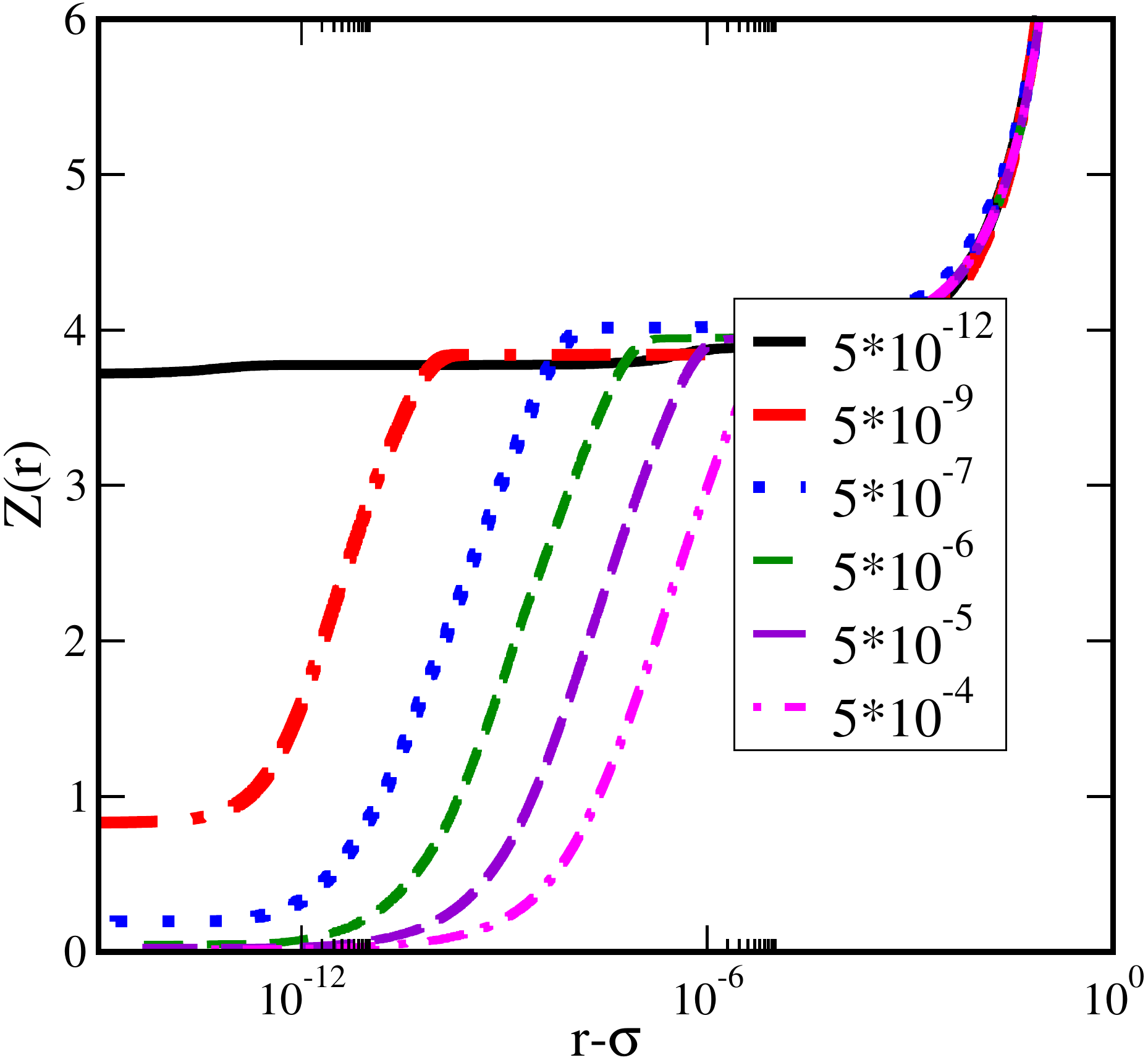}

\caption{\label{figSI2} {\bf (a)} Radial distribution function for
  $\phi = 0.61$ for different values of strain step. The distance at
  which a deprature from the power law in $g(r)$ is seen depends on
  the strain step used in shearing.  
  Cumulative distribution of neighbor number shown for different
  $d\gamma$ for {\bf (b)} $\phi = 0.61$, {\bf (c)} $\phi = 0.58$, and {\bf (d)} $\phi = 0.54$. The plateau value is the same for all $d \gamma$ values to a very good degree. In the limit of $d\gamma
  \rightarrow 0$, the plateau in $Z(r)$ extends to $r = \sigma$, indicating that the unique contact number in that limit is reliably obtained by considering the plateau value when performing finite $d\gamma$ shearing.}
\end{figure}

In our analysis, we determine the contact number geometrically in the
first instance, from the pair correlation function $g(r)$. As seen in
Fig. \ref{figSI2}(a), the pair correlation function exhibits a deviation from the
power law form at small distances (below $r - \sigma = 10^{-5}$). We
identify the contact neighbors as those at distances $r - \sigma <
tol$ where $tol$ is the value where the near contact $g(r)$ deviates
from the power law behavior. We attribute the fact that these contact
distances are not exactly equal to $\sigma$ to the finite precision of
the procedure used to generate the configurations. With that premise,
the use of a tolerance $tol$ is a reasonable procedure to identify
contact neighbors, but such neighbors will not necessarily be {\it in
  contact}, {\it i. e.} exert a finite force on each other. This
limitation can be overcome by compressing the sphere together such
that all spheres within $\sigma + tol$ come in to contact with each
other. We use such a procedure in our work when evaluating forces and
in performing frictional simulations. Here, we validate these
procedures.  In order to do so, we first show that the deviation from
the power law occurs for distance values that depend systematically on
the precision of the numerical procedure used. In the case of the
athermal quasistatic procedure we use, the limit to precision arises
from the finite strain steps $d \gamma$ used. We next show that in the
limit of vanishing $d \gamma$, the average number of contacts, defined
as pairs of spheres exerting forces on each other, becomes equal to
the average number of {\it geometric} neighbors we identify from the
$g(r)$. 

We generate steady state sheared configurations for different values
of the strain step $d\gamma$, by starting from the steady state
configurations with $d\gamma = 5 \times 10^{-5}$, and progressively
decreasing $d\gamma$, and obtaining the corresponding steady state
configurations. As shown from Fig. \ref{figSI2} (a) the location of the
departure from the power law ($tol$) depends on the $d\gamma$ value.
In Fig. \ref{figSI2} (b) - (d) we show the cumulative distribution of
neighbors, $Z(r)$, defined as the number of neighbors of a given
sphere within a distance $r$. The behavior of $Z(r)$ for different
values of $d\gamma$ shows clearly that in the limit of $d\gamma
\rightarrow 0$, the plateau value of $Z(r)$, which corresponds to
counting neighbors within $\sigma + tol$ with $tol$ identified from the
$g(r)$, becomes exact as the number of contact neighbors. Thus, for
finite strain step simulations, (a) counting the {\it geometric}
neighbors defined from $g(r)$ as above as contact neighbors, and (b)
compressing the configurations to induce contact between neighbors at
distances below $\sigma + tol$ when forces and dynamics are studied,
are both seen to be reasonble. We use these procedures, but to
minimize the approximation to the AQS procedure as much as possible,
consider the case of $d \gamma = 5 \times 10^{-12}$ when evaluating
forces, and performing frictional simulations.

\section{Lower limit for shearing jamming in the presence of friction}

Here, we provide details of these procedures we use to identify the
lowest friction coefficient for which sheared steady state
configurtions jam, and thereby, how we identify the lower density
limit to frictional jamming. 

We used the discrete element method to simulate frictional interactions between the spheres. A linear-spring dashpot model is used with forces between particles in contact given by 

\begin{equation}
\vec{F} = (\kappa_n \delta \vec{n_{ij}} - m_{eff} \zeta_n \vec{v_n}) -  (\kappa_t \Delta s_{t} + m_{eff} \zeta_t \vec{v_t}). 
\end{equation} 
The first term in the normal force between the two particles and the
second term is the tangential force. $\kappa_n$ and $\kappa_t$ is the
elastic constant for normal and tangential displacements. $\zeta_n$
and $\zeta_t$ are damping constants for the normal and tangential
contacts. $m_{eff}$ is the effective mass of the spheres. $\Delta s_t$
is the tangential displacement vector between two spheres from the
contact point.  $n_{ij}$ is the unit vector along the line connecting
the centers of the two particles and $\delta \vec{n_{ij}}$ is the
  normal displacement of the spheres (towards each other) from the
  contact position.  $v_n$ and $v_t$ are the normal and tangential
  components of the relative velocity of the two particles. The maximum value of the tangential force $F_t$ follows the Coulomb criterion, $F_t \le \mu F_n$, where $F_n$ is the normal force, and $\mu$ is the friction coefficient. \newtext{We also include solvent friction in some cases, through a frictional force term $\vec{F}_{solvent} = -\eta \vec{v}$.}

\begin{figure}[h!]
\includegraphics[scale=0.4,angle=0]{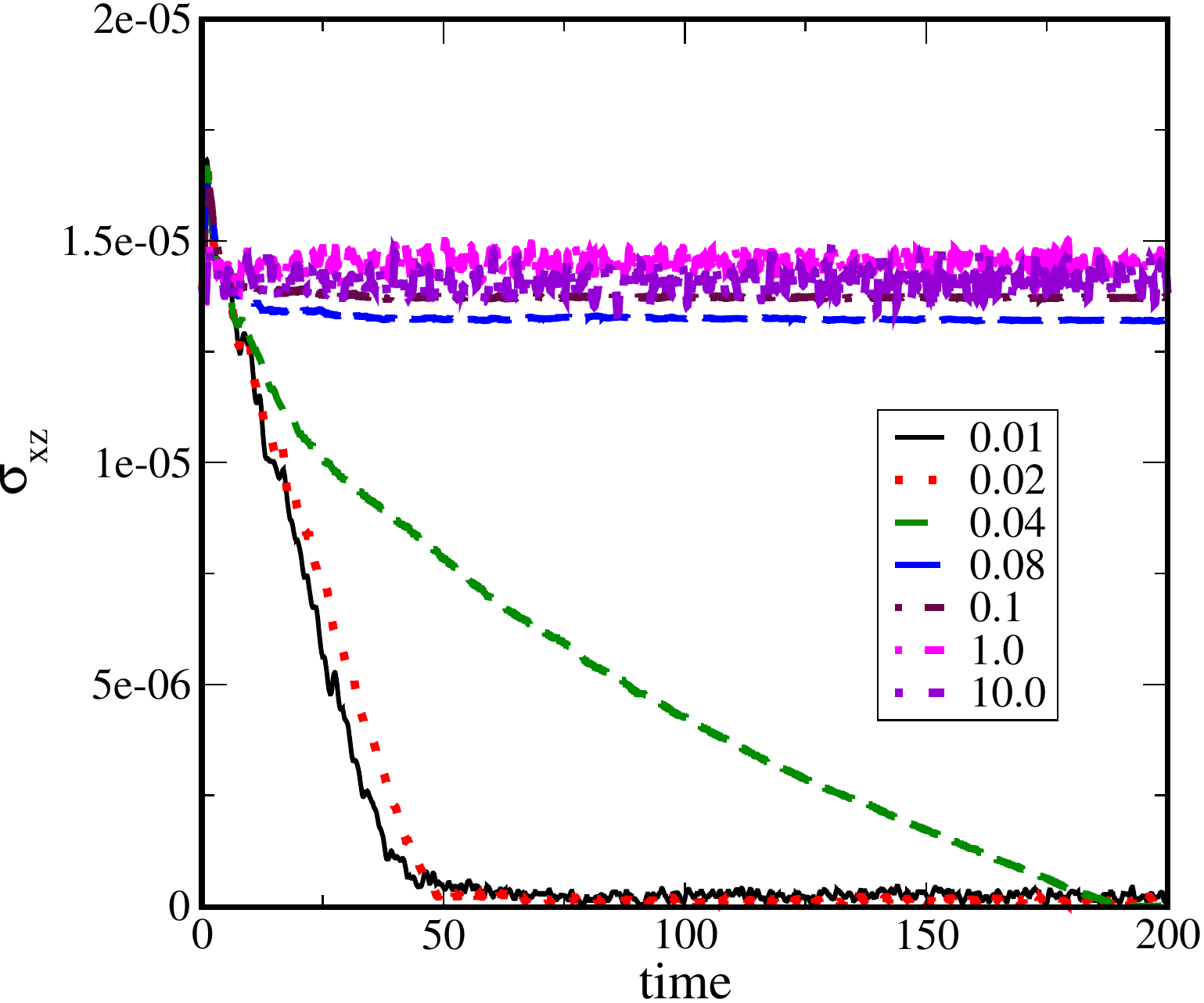}
\hspace{1cm}
\includegraphics[scale=0.4,angle=0]{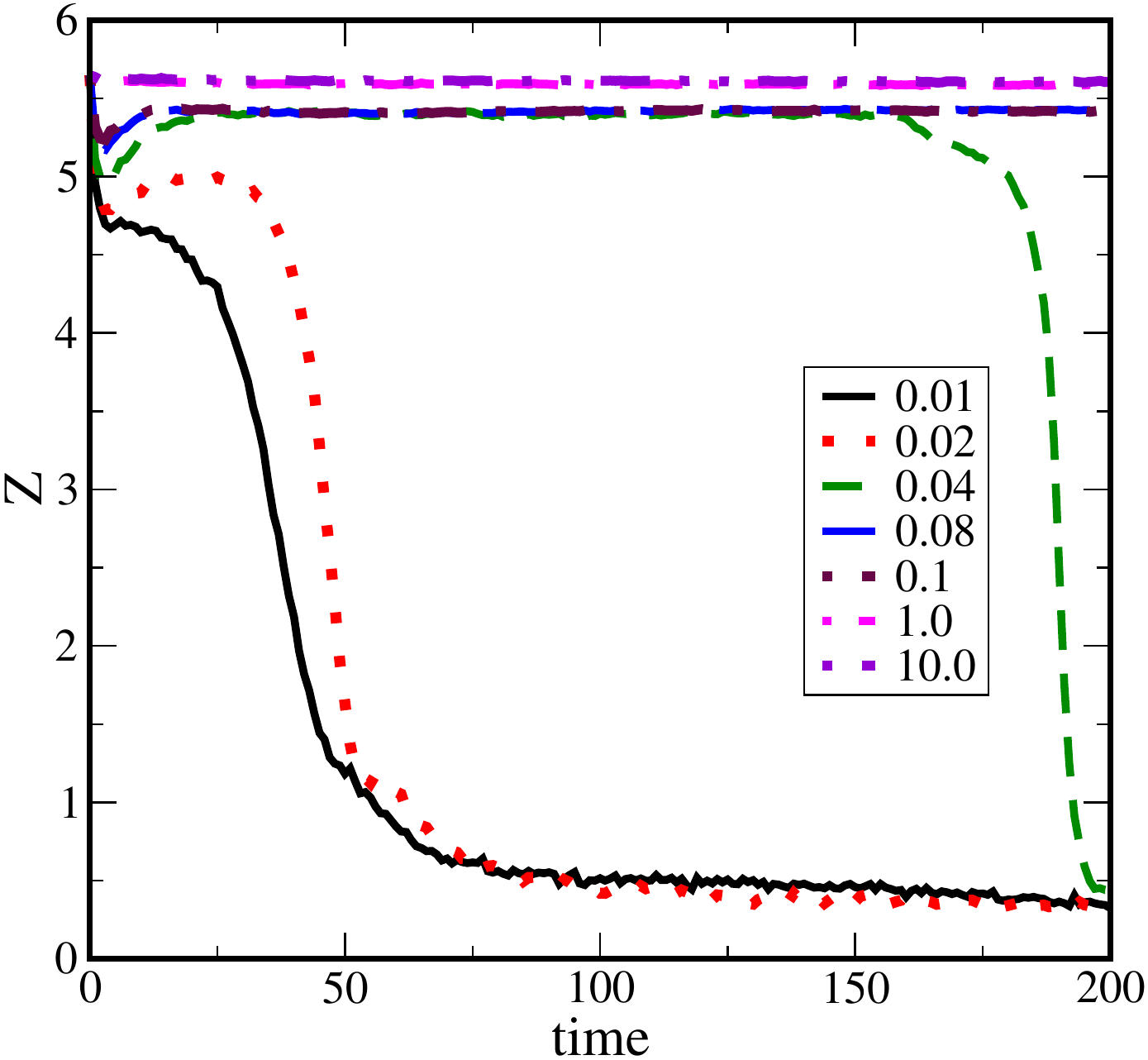}

\caption{\label{figSI3} For $\phi = 0.627$, {\bf (a)} stress, and
  {\bf (b)} contact number, as function of time for the sheared
  configurations after an initial strain of $\Delta \gamma =
  5\times10^{-5}$ for different friction coefficients. For very low
  friction coefficient, the shear stress relaxed to zero indicating
  that the structure is jammed, whereas the stress remains finite for
  large friction coefficients indicating that the structure is
  jammed. Similarly, the contact number decays to zero for structures
  that are not stable, while remaining close to the initial value for
  jammed structures.}
\end{figure}

\begin{figure}[h]
\includegraphics[scale=0.4,angle=0]{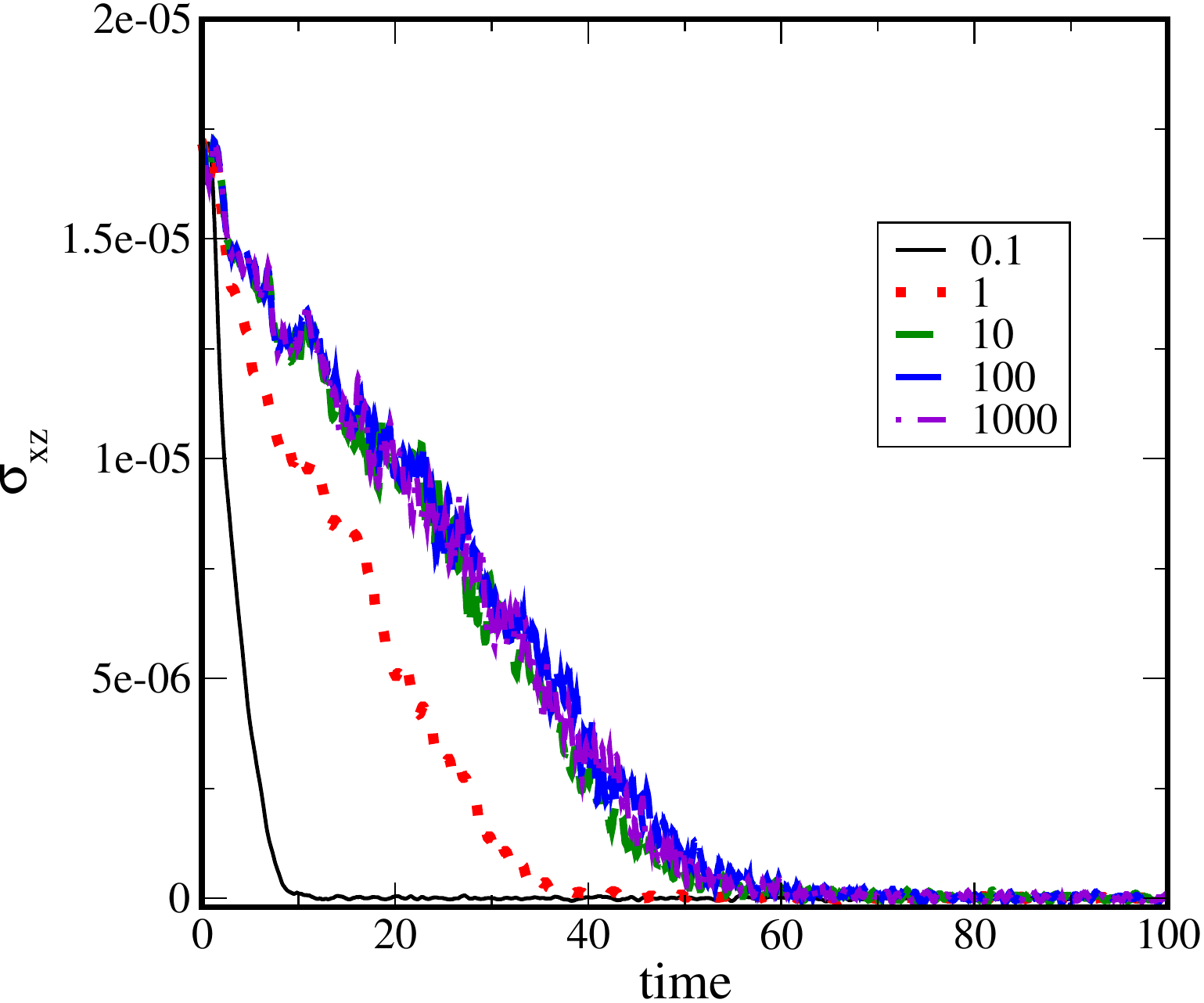}
\hspace{1cm}
\includegraphics[scale=0.4,angle=0]{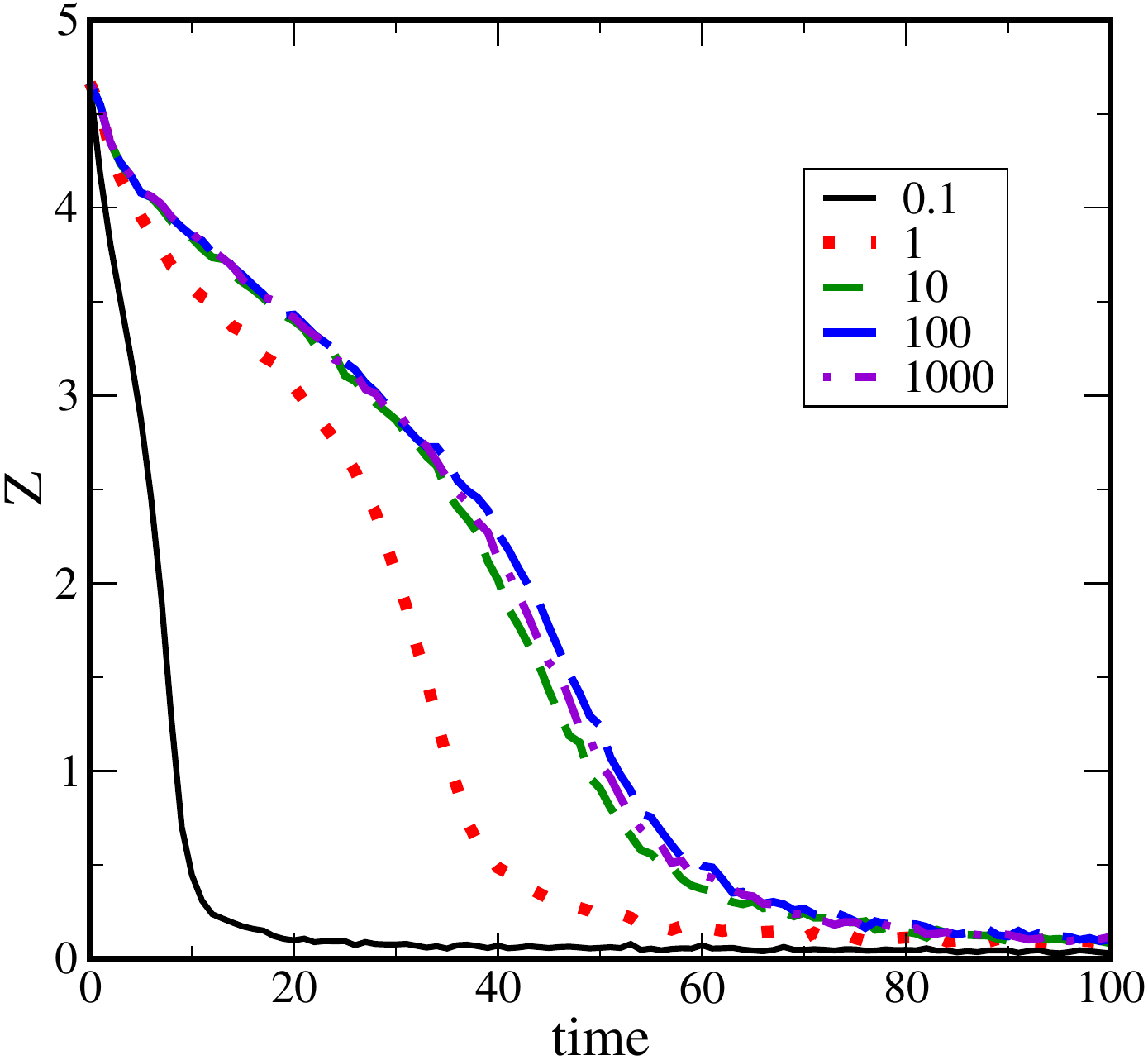}
\caption{\label{figSI4} For $\phi = 0.58$, {\bf (a)} stress, and
  {\bf (b)} contact number, as function of time for the sheared
  configurations after an initial strain of $\Delta \gamma =
  5\times10^{-5}$ for different friction coefficients. Neither the stress nor contact number remain finite for any friction coeffficients, indicating that the structures are not stable in the studied range of friction. }
\end{figure}

\begin{figure}[h]
\includegraphics[scale=0.4,angle=0]{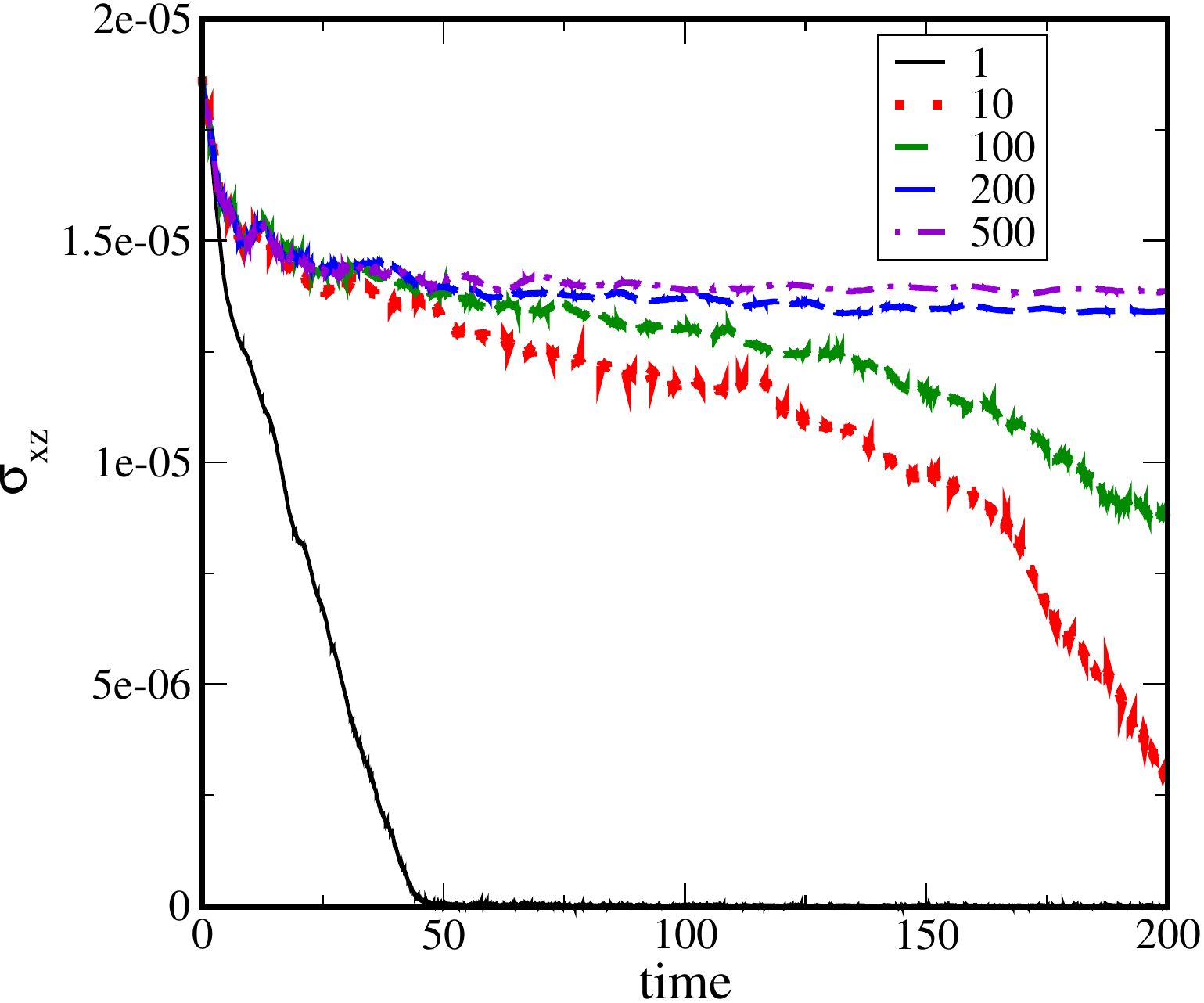}
\hspace{1cm}
\includegraphics[scale=0.4,angle=0]{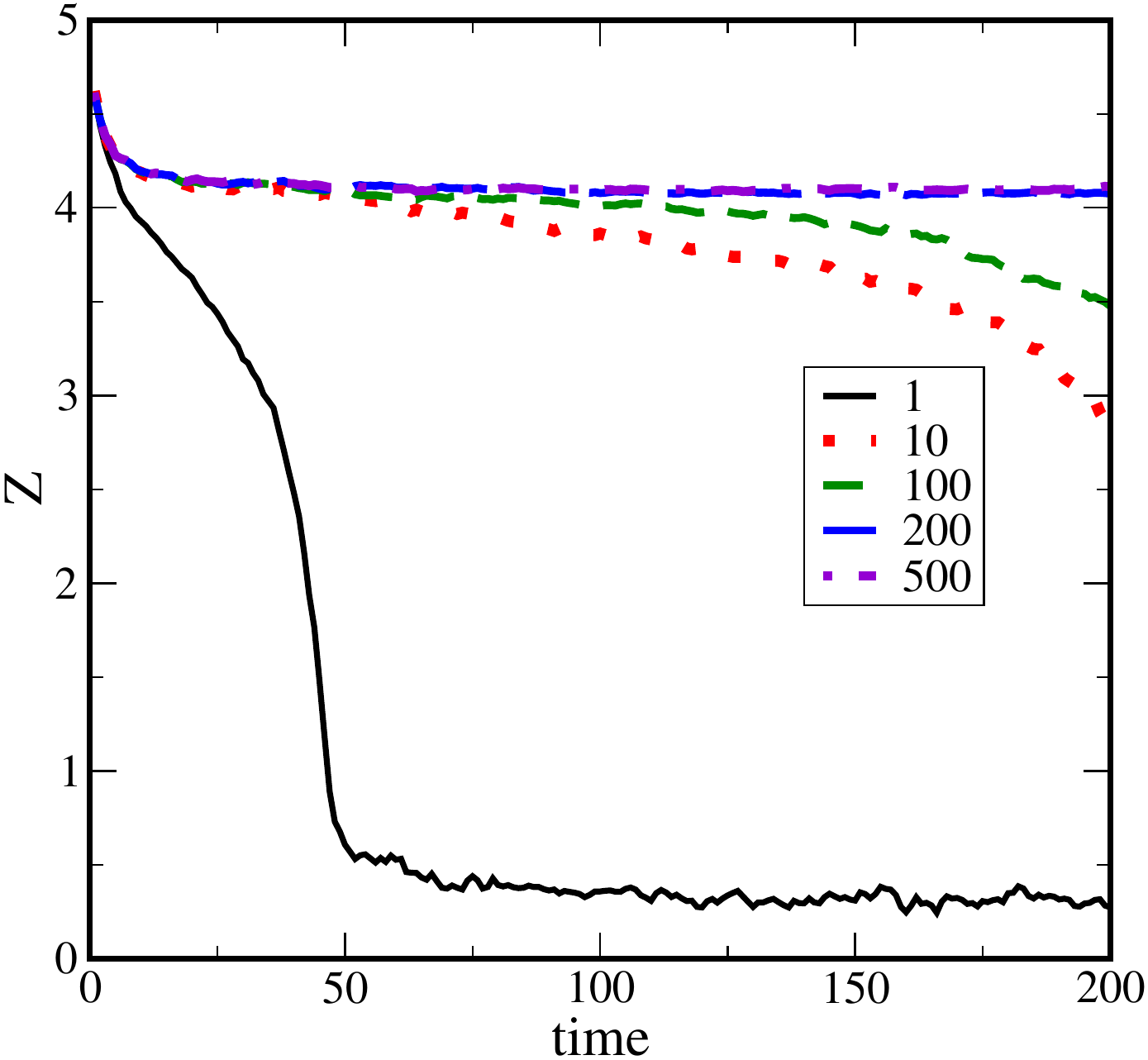}
\caption{\label{figSI5} For $\phi = 0.58$, {\bf (a)} stress, and
  {\bf (b)} contact number, as function of time for the sheared
  configurations after an initial strain of $\Delta \gamma =
  5\times10^{-5}$ for different friction coefficients with normal damping constant $\zeta_n = 3$. The stress and contact number remain finite for high friction coeffficients, indicating that the structures can be stabilized by adding small 
amount of damping. }
\end{figure}

The input parameters used in our analysis is $\kappa_n = \kappa_t = 2$,
$\zeta_n = 0,3,30$ and $\zeta_t = \frac{1}{2} \zeta_n$. We
use sheared steady state structures, compressed by $tol$ to form all the
contacts that are counted (see previous section) and strained by a
variable step $\Delta \gamma$, as initial configurations. Stability of the
sheared structures are tested by applying strain steps $ \Delta \gamma$ in
the range $0$-$10^{-2}$ and allowing the system to relax. By varying
the friction coefficient, we identify, for each density, the friction
coefficients for which the initial structure remains stable.  Fig. \ref{figSI3} shows that for $\phi = 0.627$, the packings have finite stress and
contact number after applying a strain step of $\Delta \gamma = 5*10^{-5}$
for a range of friction coefficients. The decay or otherwise of the
shear stress, and the contact number provide clear and consistent
criteria for judging whether the sytem remains jammed for a given
friction coefficient $\mu$. For $\phi = 0.58$, shown in Fig. \ref{figSI4}, the
stresses and contact numbers decay to zero for all the friction
coefficients studied, and thus in this range of $\mu$, the system does
not jam. It appears that the limit value in density for any friction
coefficient will also depend on other parameters, namely the damping
coefficients. Fig. \ref{figSI5} shows the same data for $\phi = 0.58$, but for a damping coefficient of $\zeta_n = 3$, which indeed shows that for finite damping, the system is jammed above a threshold friction coefficient. 

\begin{figure}[h]
\includegraphics[scale=0.4,angle=0]{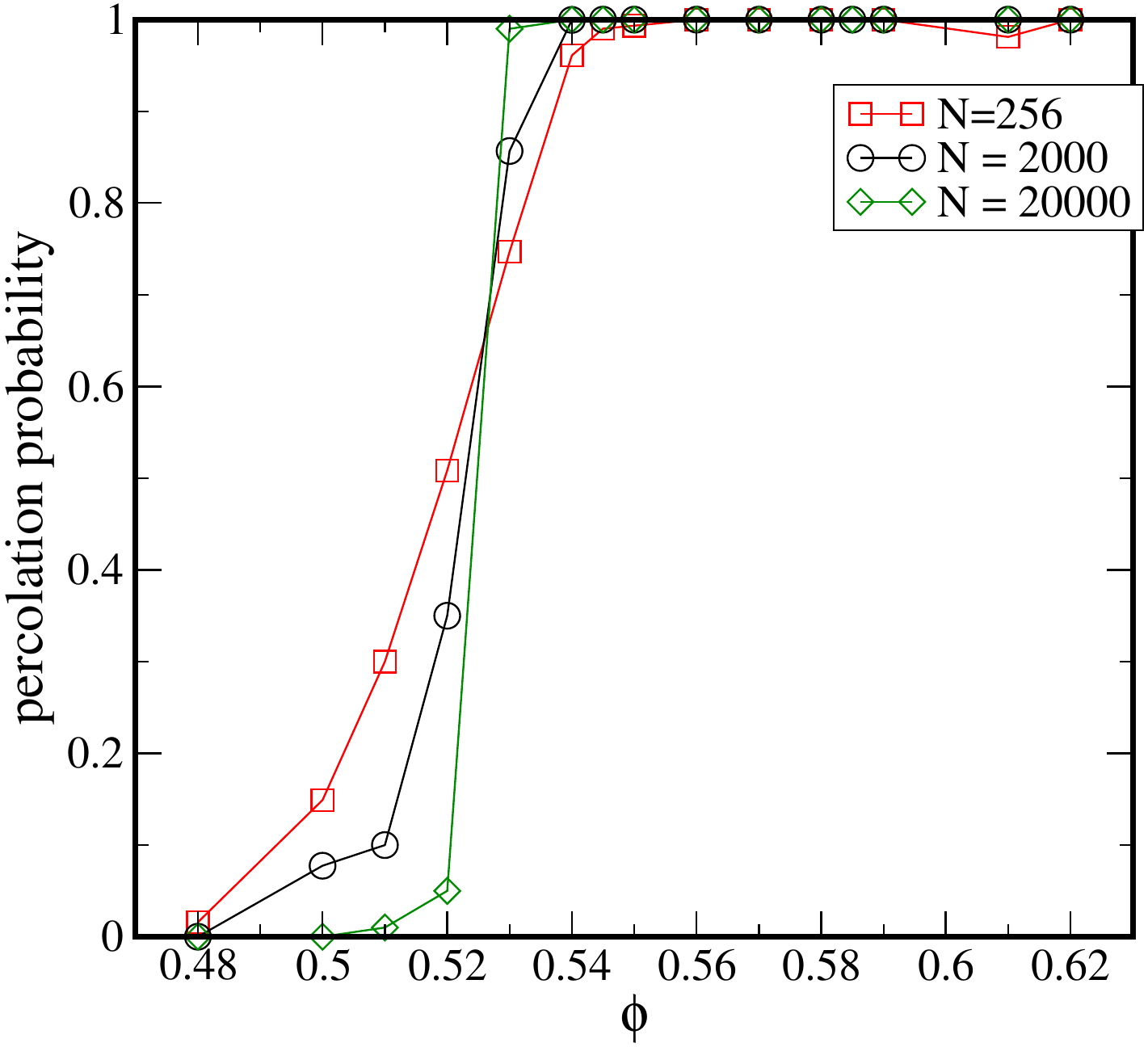}
\caption{\label{figSI6} \newtext{Percolation of spheres with $D+1$ contacts, such that all the neighbors are not on the same half space with respect to the central sphere.}}
\end{figure}

\newtext{In Fig. \ref{figSI6}, we show the percolation of spheres with $D + 1$
  contacts, such that the contacts are not all in the same half space
  with respect to the central sphere, for different system sizes. It
  is seen that the percolation transition occurs at density $\phi =
  0.53$. This threshold is a little lower than $\phi = 0.55$ where we
  see other indications of a change in behavior, and its significance
  remains to be understood better.}

\begin{figure}[h]
\includegraphics[scale=0.4,angle=0]{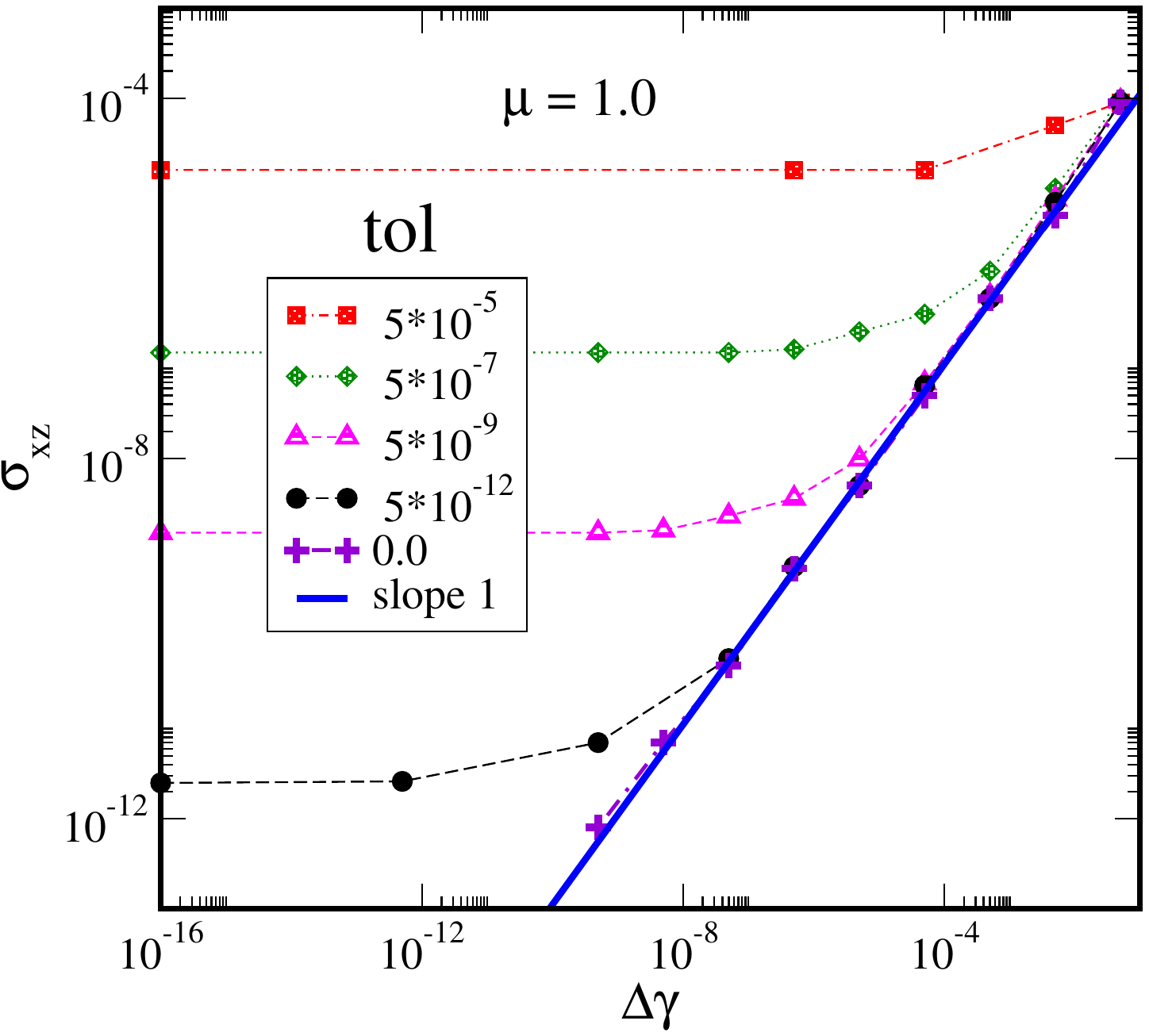}
\caption{\label{figSI7} Stress as a function of applied strain
  $\Delta \gamma$ for $\phi = 0.61$ and $\mu = 1.0$, shown for
  configurations obtained using different strain steps $d\gamma$. The
  initial sheared configurations are compressed by an amount $tol$ that
  depends on $d\gamma$ to include all the contacts before the strain
  $\Delta \gamma$ is applied. For $\Delta \gamma > tol$, the shear stress is
  proportional to strain as expected for an elastic solid.}
\end{figure}

Finally, we consider the stress the jammed stuctures should experience
at the end of the relaxation. For jammed configurations, one should
expect a linear stress-strain dependence, resulting in a finite shear
modulus from the slope of the stress-strain curve. However, since the
frictional dynamics is initiated after a compression that is
determined by the strain step $d \gamma$, the jammed packings exhibit
an initial stress value that is independent of the strain $\Delta
\gamma$. To demostrate that this is an artefact of finite precision,
we consider steady state configurations obtained with different $d
\gamma$, and consider the final stress-strain curves {\it vs.} $\Delta
\gamma$. We see, in Fig. \ref{figSI7}, that as $\Delta \gamma$ exceeds $d \gamma$, we obtain
a linear stress-strain relation, from which a modulus may be
extracted.

\begin{thebibliography}{100}




\bibitem{liu-nagel-2010} Liu, A. J. \& Nagel, S. R. \textrm{The Jamming Transition and the Marginally Stable Solid}. \textit{Annu. Rev. Condens. Matter Phys.} \textbf{1} 347-369 (2010).

\bibitem{bernal-mason-1960} Bernal, J. D. and Mason, J. {Packing of Spheres: Co-ordination of Randomly Packed Spheres}, \textit{Nature}  \textbf{188}, 910 - 911 (1960).

\bibitem{scott-1960} Scott, G. D. \textrm{Packing of Spheres: Packing of Equal Spheres} \textit{Nature} \textbf{188}, 908-909 (1960)

\bibitem{torquatorev} Torquato, S. and Stillinger, F. H. \textrm{Jammed hard-particle packings: From Kepler to Bernal and beyond}  {\it Rev. Mod. Phys.} {\bf  82}, 2633Ð2672 (2010).

\bibitem{pinaki-2010} Chaudhuri, P.,  Berthier, L., and Sastry, S. \textrm{Jamming Transitions in Amorphous Packings of Frictionless Spheres Occur over a Continuous Range of Volume Fractions}, \textit{Phys. Rev. Lett.} \textbf{104}, 165701 (2010).

\bibitem{ohern-2002} O'Hern, C. S., Langer, S. A., Liu, A. J. and Nagel, S. R., \textrm{Random Packings of Frictionless Particles}, \textit{Phys. Rev. Lett.} \textbf{88} 075507 (2002). 

 
\bibitem{makse-2008}Song, C., Wang, P. \& Makse, H. A. \textrm{A phase diagram for jammed matter}. \textit{Nature}
\textbf{453,} 629-632(2008). 

\bibitem{silbert-2010} Silbert, L. E. \textrm{Jamming of frictional spheres and random loose packing}, \textit{Soft Matter} 
\textbf{6,} 2918-2924(2010).

\bibitem{bi-2011} Bi, D., Zhang, J., Chakraborty, B. \& Behringer, R. p. \textrm{Jamming by shear}. \textit{Nature} \textbf{480,} 
355-358(2011).

\bibitem{otsuki-2011} Otsuki, M. and Hayakawa, H. \textrm{Critical scaling near jamming transition for frictional granular particles}. \textit{Phys. Rev. E}, \textbf{83} 051301 (2011).


\bibitem{ciamarra-2011} Ciamarra, P., Pastore, R., Nicodemi, M. and Coniglio, A. \textrm{Jamming phase diagram for frictional particles}. \textit{Phys. Rev. E}, \textbf{84} 041308 (2011). 



\bibitem{shen-2012} Shen, T., O'Hern, C. S. and Shattuck, M. D. \textrm{Contact percolation transition in athermal particulate systems} \textbf{85} 011308 (2012).                                                    

\bibitem{claus-2009} Heussinger, C. and Barrat, J.-L. \textrm{Jamming Transition as Probed by Quasistatic Shear Flow} \textit{Phys. Rev. Lett.} \textbf{102} 218303 (2009).

\bibitem{onoda-1990}Onoda, G. Y. \& Liniger, E. G. \textrm{Random loose packings of uniform spheres and the dilatancy onset}.
\textit{Phys. Rev. Lett.} \textbf{64,} 2727-2730 (1990).

\bibitem{bertrand} \newtext{Bertrand, T. {\it et al} \textrm{Protocol dependence of the jamming transition}, arXiv:1506.05041.}

\bibitem{kumar} \newtext{Kumar, N. and Luding, S. \textrm{ Memory of jamming - multiscale flow in soft and granular matter.} arXiv:1407.6167.} 

\bibitem{ohern-2003} O'Hern, C. S., Silbert, L. E., Liu, A. J \& Nagel, S. R. \textrm{Jamming at zero temperature and zero
applied stress: The epitome of disorder}. \textit{Phys. Rev. E.} \textbf{68,} 011306(2003). 

\bibitem{donev-2005}Donev, A., Torquato, S., \& Stillinger, F. H. \textrm{Pair correlation function characteristics of nearly 
jammed disordered and ordered hard-sphere packings}. \textit{Phys. Rev. E.} \textbf{71 ,} 011105(2005).

\bibitem{maiti-2014} Maiti, M. and Sastry, S. \textrm{Free volume distribution of nearly jammed hard sphere packings}. 
\textit{J. Chem. Phys.} \textbf{141,} 044510 (2014). 

\bibitem{maiti-unpub} \newtext{Maiti. M., Vinutha, H. A., Heussinger, C. and Sastry, S. \textrm{Free Volume under Shear}, {\it J. Chem. Phys.} (in press). }

\bibitem{ohern-2001} O'Hern, C. S., Langer, S. A., Liu, A. J. and Nagel, S. R., \textrm{Force Distributions near Jamming and Glass Transitions}, \textit{Phys. Rev. Lett.} \textbf{86} 111-114 (2001). 

\bibitem{wyart-2012} Wyart, M. \textrm{Marginal Stability Constrains Force and Pair Distributions at Random Close Packing} \textit{Phys. Rev. Lett.} \textbf{109} 125502 (2012).

\bibitem{lerner-2013} Lerner, E., D\"ruing, G. and Wyart, M. \textrm{Low-energy non-linear excitations in sphere packings} \textit{Soft Matter} \textbf{9}, 8252-8263 (2013).

\bibitem{LS}Lubachevsky, B. D. \& Stillinger, F. H. \textrm{Geometric properties of random disk packings}.
\textit{J. Stat. Phys.} \textbf{60,} 561-583(1990).


\bibitem{speedy-1998} Speedy, R. J. \textrm{The hard sphere glass transition}. \textit{Mol. Phys.} \textbf{95,} 169-178(1998).


\bibitem{brown-2009}Brown, E. \& Jaeger, H. M. \textrm{Dynamic jamming point for shear thickening Suspensions}. \textit{Phys. Rev. Lett.} \textbf{103,} 086001(2009).


\bibitem{cates} Cates, M. E., Wittmer, J. P., Bouchaud, J.-P. and Claudin, P. \textrm{Jamming, Force Chains, and Fragile Matter}. \textit{Phys. Rev. Lett.} \textbf{81,} 1841 (1998).

\bibitem{henkes-2010}  Henkes, S., van Hecke, M. and van Saarloos, V. \textrm{Critical jamming of frictional grains in the generalized isostaticity picture} \textit{Euro Phys. Lett.}  \textbf{90} 14003 (2010).  

\bibitem{lammps} Plimpton, S. \textrm{Fast Parallel Algorithms for Short-Range Molecular Dynamics}. 
\textit{J. Comput. Phys.} \textbf{117} 1Ð19 (1995). 

\bibitem{cundall} Cundall, P. A. \& Strack, O. D. L. \textrm {A discrete numerical model for granular assemblies}.\textit{Geotechnique} \textbf{29}, 47–65 (1979). 

\end{thebibliography}
\end{document}